\documentclass[hydrogen,article,accept,pdftex,moreauthors]{Definitions/mdpi} 

\firstpage{1} 
\pubvolume{1}
\issuenum{1}
\articlenumber{0}
\pubyear{2022}
\copyrightyear{2022}
\datereceived{} 
\dateaccepted{} 
\datepublished{} 
\hreflink{https://doi.org/} 
\usepackage{amssymb,amsmath,mathrsfs,stmaryrd,mathtools,amsfonts,graphicx,url,booktabs,nicefrac,comment,microtype,tikz-cd,lingmacros,nccmath,tree-dvips, bbm, bm, etoolbox,algorithm,caption,subcaption,color,enumitem, physics, lineno,setspace,multirow}

\Title{Distributional trends in the generation and end-use sector of low-carbon hydrogen plants}

\TitleCitation{Distributional trends in the generation and end-use sector of low-carbon hydrogen plants}
\Author{Nick James $^{1,\ddagger}$\orcidA{}, Max Menzies $^{2,\ddagger}$\orcidB{}}
\AuthorNames{Nick James and Max Menzies}
\AuthorCitation{James, N.; Menzies, M.}
\address{%
$^{1}$ \quad School of Mathematics and Statistics, University of Melbourne, Victoria 3010, Australia; nick.james@unimelb.edu.au \\
$^{2}$ \quad Beijing Institute of Mathematical Sciences and Applications, 
Tsinghua University, Beijing 101408, China; max.menzies@alumni.harvard.edu
}
\corres{Correspondence: max.menzies@alumni.harvard.edu (M.M.)}
\secondnote{These authors contributed equally to this work.}
\abstract{
This paper uses established and recently introduced methods from the applied mathematics and statistics literature to study trends in the end-use sector and capacity of low-carbon hydrogen projects in recent and upcoming decades. First, we examine distributions in plants over time for various end-use sectors and classify them according to metric discrepancy, observing clear similarity across all industry sectors. Next, we compare the distribution of usage sectors among different continents and examine the changes in sector distribution over time. Finally, we judiciously apply several regression models to analyse the association between various predictors and the capacity of global hydrogen projects. Across our experiments, we see a welcome exponential growth in the capacity of zero-carbon hydrogen plants and significant growth of new and planned hydrogen plants in the 2020's across every sector.
}


\begin{document}

\section{Introduction}
\label{sec:intro}

Hydrogen has great potential as an alternative fuel source to existing fossil fuels and may play a role in the world's coordinated attempt to reach net-zero carbon emissions during this century. Hydrogen production has already been used for numerous purposes, including production of liquid fuels, water, and space heating; direct injection into the gas grid; transport; and~other industrial applications. Unlike petrol or natural gas, the~combustion of hydrogen does not produce any greenhouse gas (GHG) emissions, which have a role in anthropogenic global warming. Hydrogen in its molecular form ($H_2$ gas) does not exist naturally on Earth; it must be synthesised via a variety of different~procedures.

The production of hydrogen is classified by colours according to its mode of preparation and associated emissions. ``Green hydrogen'' refers to production techniques that do not generate any GHG emissions, the~most desirable outcome from a sustainability perspective. Typically, green hydrogen plants use renewable sources of energy \mbox{(such as solar)} to extract hydrogen via the electrolysis of water; the acquired hydrogen may then be stored and subsequently used for numerous applications. Black, brown, and grey hydrogen refer to production techniques that use black coal, brown coal, and natural gas, respectively, and are increasingly obsolete and costly energy sources with considerable amounts of emissions and harmful byproducts~\cite{coal_environment}. Blue hydrogen is a middle ground, defined as the production of hydrogen using fossil fuels followed by carbon capture and storage (CSS). This is not a truly zero-emissions process, as~only a proportion of all generated GHGs can be captured, and~harmful byproducts may~remain.

With the increasing global interest in alternative energy sources in recent years, a~great deal of research has focused on the viability and underlying science of hydrogen production. Early research into hydrogen fuel cells dates back several decades~\cite{doenitz_hydrogen_1980,wendt_hydrogen_1991,khaselev_high-efficiency_2001}. Since then, research on hydrogen production has been substantial: \cite{ursua_hydrogen_2012,chi_water_2018} each provided a review article on different electrolysis technologies,~\cite{zeng_recent_2010} detailed alkaline electrolysis (ALK),\mbox{ and~\cite{barbir_pem_2005,grigoriev_pure_2006,bessarabov_pem_2016}} described successive advances in proton exchange membrane electrolysis (PEM). \mbox{Reference \cite{guo_comparison_2019}} compared and contrasted ALK and PEM in~detail.

More recent research has explored sophisticated means for integrating different renewable energy sources. In~\cite{yilmaz_thermodynamic_2014}, the authors explored the use of geothermic energy to power  electrolysis, and~\cite{rozendal_principle_2006} investigated the use of micro-organisms (biocatalysed electrolysis). Then,~\cite{call_hydrogen_2008} discussed further advances in such microbial electrolysis cells. More recently,~\cite{Rai2022} described how electronic waste may be used to generate metallic components for a process called \mbox{``chemical looping~reforming''.}

There have also been many technological advances to enhance the efficiency of hydrogen production, both novel and incremental. In~\cite{wang_intensification_2014}, the authors examined numerous means to reduce energy consumption during electrolysis. In~\cite{chakik_effect_2017}, the authors compared the efficiency of different electrodes, with~further advances made by~\cite{Wu2022}. In~particular,~\cite{Ikeda2022} analysed electrode overpotential during ALK electrolysis production, whereas~\cite{zhang_evaluation_2010} explored optimal configurations of electrolysis under different conditions. In~\cite{Shit2022}, the authors investigated state-of-the-art electrocatalysts, and~\cite{Tong2022} analysed the use of hybrid structures for increasingly efficient water electrolysis, combining both morphological features and electrochemical properties, whereas various researchers have studied other cutting-edge catalysts~\cite{Liang2022,Khasanah2022,Liu2022_hydrogen,UlateKolitsky2022,He2022_hydrogen}. Numerous advances in blue hydrogen plants have also been made, including pyrolysis of plastic~\cite{Soowski2022} and catalytic decomposition of methane~\cite{Shelepova2022} and other hydrocarbons~\cite{Vedyagin2021}. Finally, numerous articles have examined different means for transportation and storage of hydrogen, an~essential component of its \mbox{widespread use~\cite{Lys2020_hydrogen,Heinemann2022,Pistidda2021,Ekhtiari2022}}.

In addition, numerous articles have taken a geopolitical focus, analysing both the policy environment and natural resources of various countries and their suitability for hydrogen production plants. Different papers have investigated the significant variability in countries' adoption of hydrogen technologies, including in the United States~\cite{lattin_transition_2007}, the~United Kingdom~\cite{park_country-dependent_2013}, China~\cite{yuan_hydrogen_2010}, South Korea~\cite{park_country-dependent_2013}, the~Philippines~\cite{collera_opportunities_2021}, Mexico~\cite{ramirez-salgado_roadmap_2004},  Morocco~\cite{touili_technical_2018}, and~across the European Union~\cite{apak_renewable_2012}. Our previous work analysed the geographic rollout of low-carbon plants across different continents~\cite{james2021_hydrogen}. {Finally, numerous authors have discussed the end-use sectors~\cite{Bridgeland2022,Saeedmanesh2018} or points of end-use~\cite{Dawood2020} hydrogen energy usage, including~\cite{Pleshivtseva2023}, who comprehensively reviewed the same low-carbon plants we \mbox{mathematically analyse.}}

Whereas the existing literature has been more technological or geopolitical in focus, our paper is a mathematical study of trends in the rollout and prevalence of low-carbon hydrogen plants (green and blue) with a focus on the end-use sector. We make use of time series analysis 
 and metrics that have been extensively applied to various fields such as epidemiology~\cite{Manchein2020,james2021_CovidIndia,Li2021_Matjaz,Blasius2020,james2021_TVO,Perc2020,Machado2020}, environmental sciences~\cite{james2022_CO2,Khan2020,james2020_LP}, finance~\cite{Drod2021_entropy,james_georg,Liu1997,Basalto2007,Wtorek2021_entropy,james_arjun,Drod2001,james2022_stagflation,Gbarowski2019,james2021_MJW}, \mbox{cryptocurrencies~\cite{Sigaki2019,Drod2020_entropy,James2021_crypto2,Drod2020,Wtorek2020}}, crime~\cite{james2022_guns,Perc2013,James2023_terrorist} and other fields~\cite{Ribeiro2012,james2021_spectral,Merritt2014,james2021_olympics,Clauset2015}. We are unaware of any instance where time series or distance analysis has been applied to the rollout of hydrogen plants over time and by the end-use sector. We study the changing usage and energy capacity of low-carbon hydrogen plants over time, with~a particular interest in the increasing potential of green hydrogen plants, which emit zero carbon. Our main finding is promising: an exponential increase in the capacity of green plants over time, regardless of usage sector, and~a dramatic closing of the gap between the capacity of green and non-green~plants.

\section{Data}
\label{sec:data}
All data analysed in this paper are drawn from the International Energy Agency (IEA) \cite{Hydrogendata} and~consist of plants built in (or projected to be built in) 1975--2043. This dataset records all low-carbon hydrogen plants, namely either green, zero emissions, or~blue, incorporating fossil fuels and carbon capture and storage (CSS), as~discussed in Section~\ref{sec:intro}. We will refer to blue hydrogen plants as ``Fossil'' throughout the manuscript to highlight this distinction. The~dataset contains four different technologies of green plants: alkaline electrolysis (ALK), proton exchange membrane electrolysis (PEM), solid oxide electrolysis cells (SOEC), and~other electrolysis. The~technologies of blue plants are coal gasification, natural gas reforming, and~oil-based processes, in each case followed by CSS. Throughout the paper, we aggregate the four green electrolysis technologies and the three blue CSS production methods to classify each plant as green or blue/fossil. We also make use of the IEA estimated zero-carbon hydrogen capacity, measured in nm$^3$/h of hydrogen for each plant. This is either quoted directly from the plant or estimated according to the stated power consumption of the plant and its technology. It is a measure of how much hydrogen the plant produces (or will produce upon completion).

The most important aspect of the dataset we analyse is the end-use sector of each plant. These are: refining (oil refining), ammonia (ammonia production), methanol \mbox{(methanol production)}, iron and steel (steelmaking and other high-temperature iron processes), other industry (other high-temperature heat industrial applications), mobility (use in vehicles), power (use in the supply of power to the electricity grid), grid injection (injection of hydrogen into the natural gas grid), CHP (combined heat and power fuel cells), domestic heat (water and space heating), biofuels (biofuel production), synfuels (synthetic liquid fuels other than methanol), CH4 grid injection (injection of synthetic methane into the gas grid), and~CH4 mobility (use of synthetic methane in vehicles). A~small number of plants have more than one end-use sector, but~the majority have only~one.

\section{Distributions of end-use sector over time}
\label{sec:CDFs}

\begin{figure}[H]
    
\begin{subfigure}[b]{0.5\textwidth}
        \includegraphics[width=\textwidth]{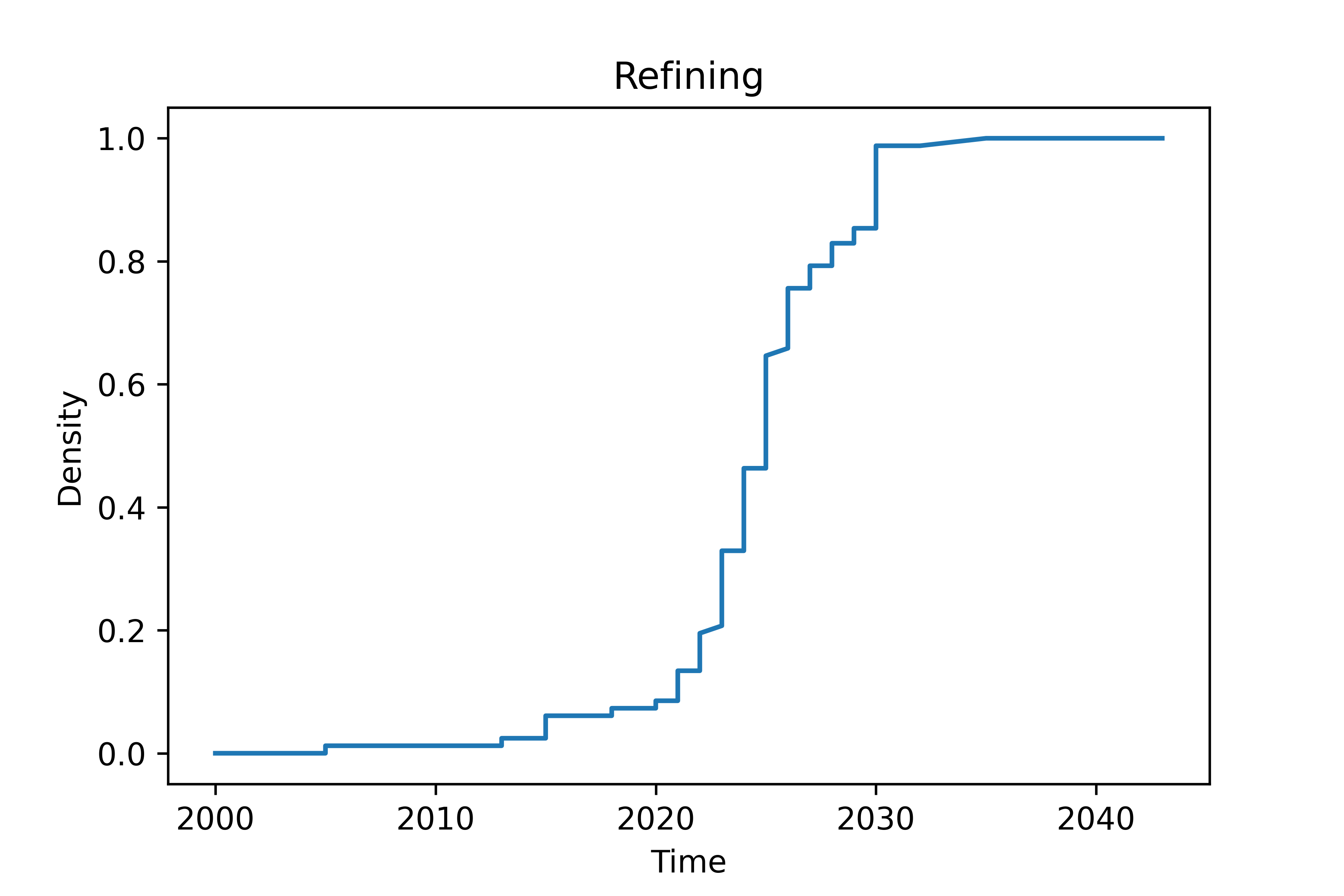}
        \caption{}
        \label{fig:CDF_Refining}
    \end{subfigure}
    \begin{subfigure}[b]{0.5\textwidth}
        \includegraphics[width=\textwidth]{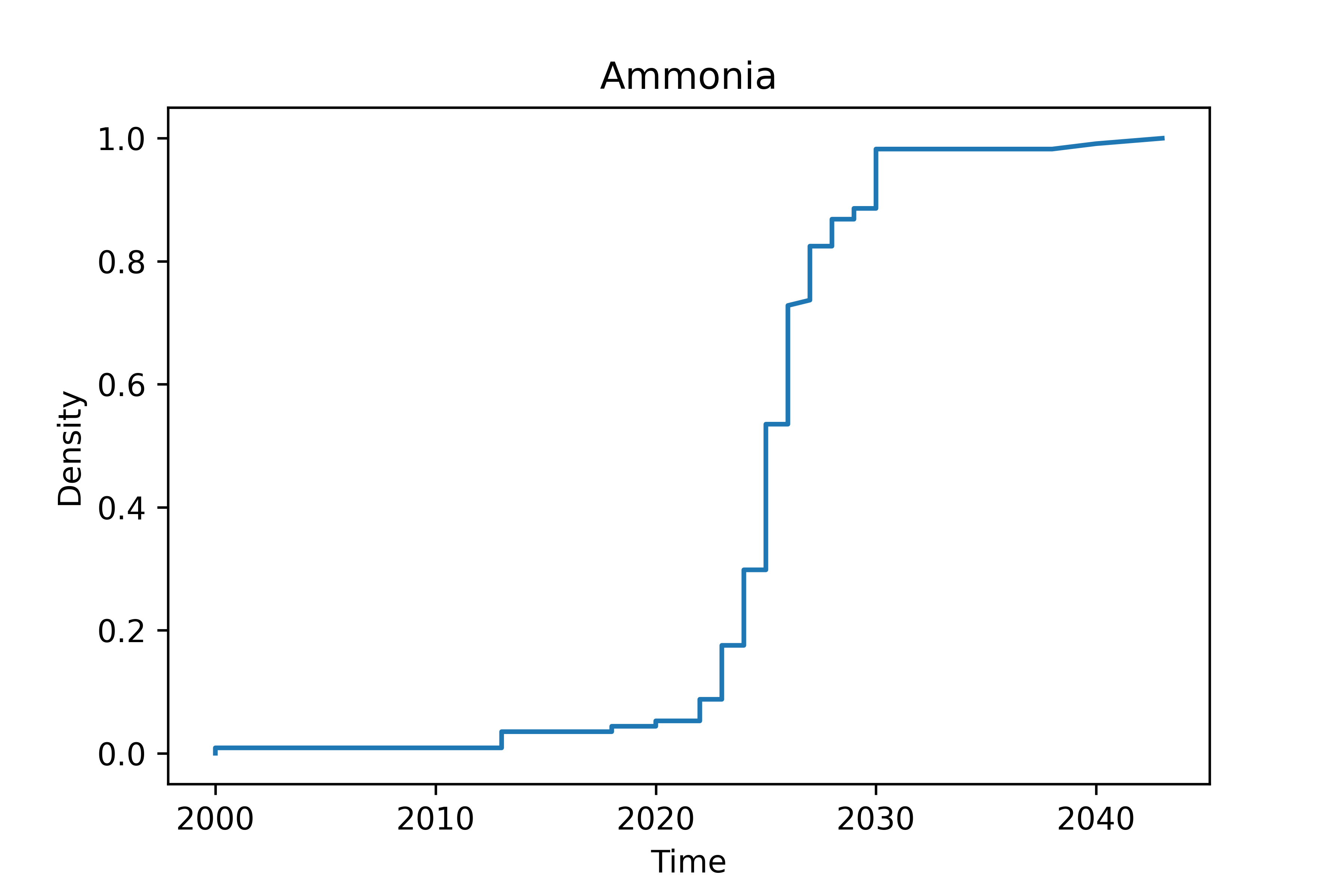}
        \caption{}
        \label{fig:CDF_Ammonia}
    \end{subfigure}
        \begin{subfigure}[b]{0.5\textwidth}
        \includegraphics[width=\textwidth]{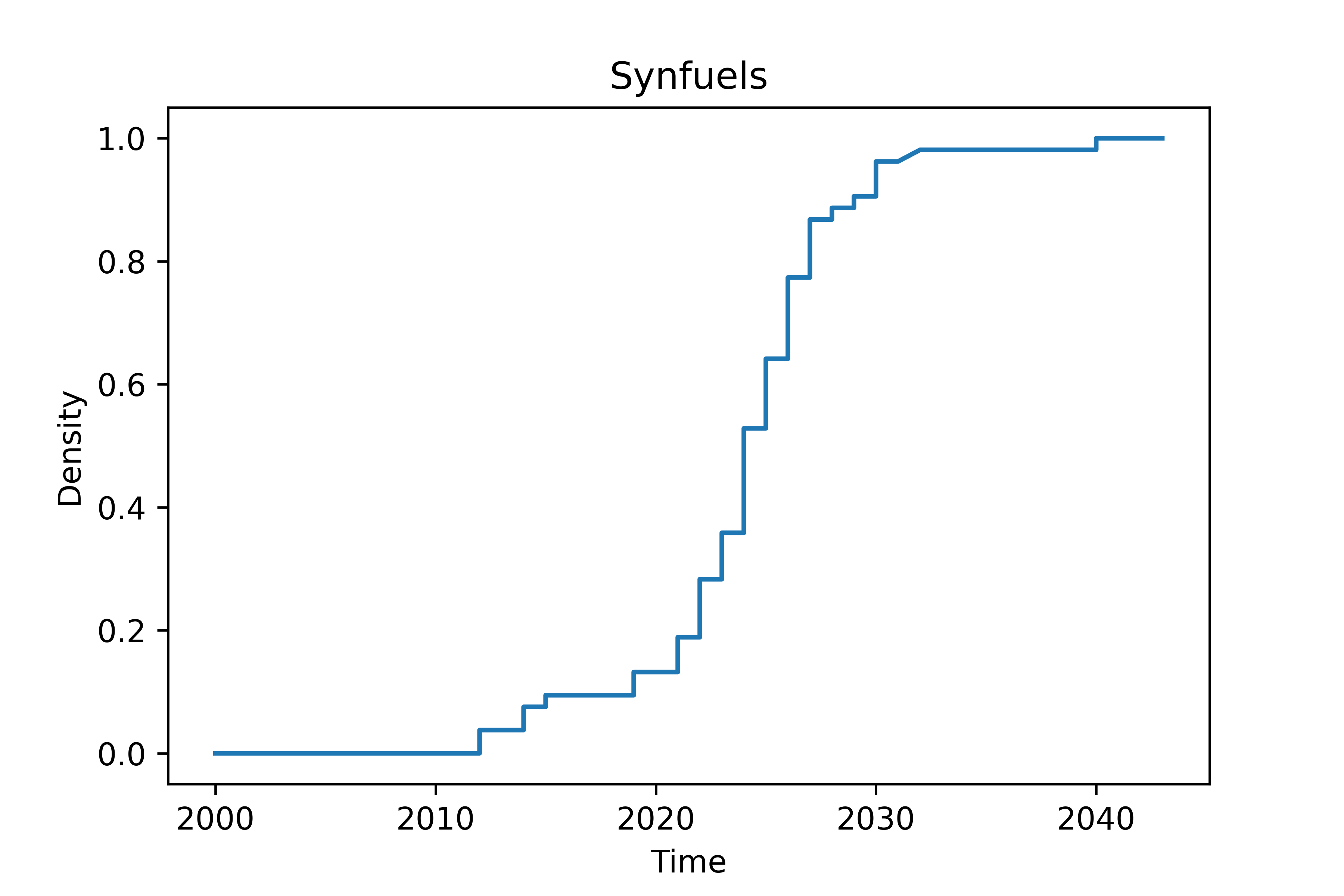}
        \caption{}
        \label{fig:CDF_Synfuels}
    \end{subfigure}
\begin{subfigure}[b]{0.5\textwidth}
        \includegraphics[width=\textwidth]{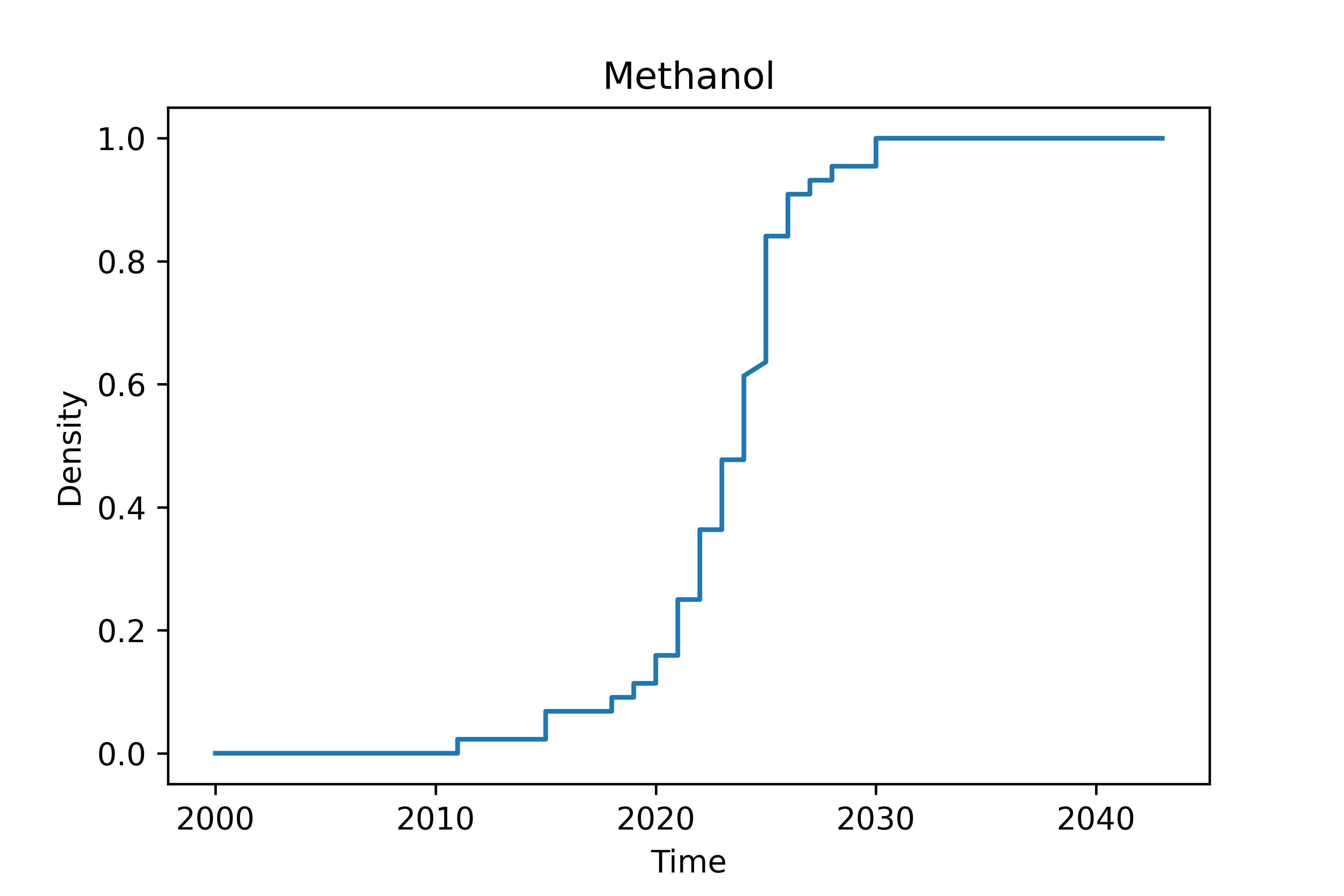}
        \caption{}
        \label{fig:CDF_Methanol}
    \end{subfigure}
        \begin{subfigure}[b]{0.5\textwidth}
        \includegraphics[width=\textwidth]{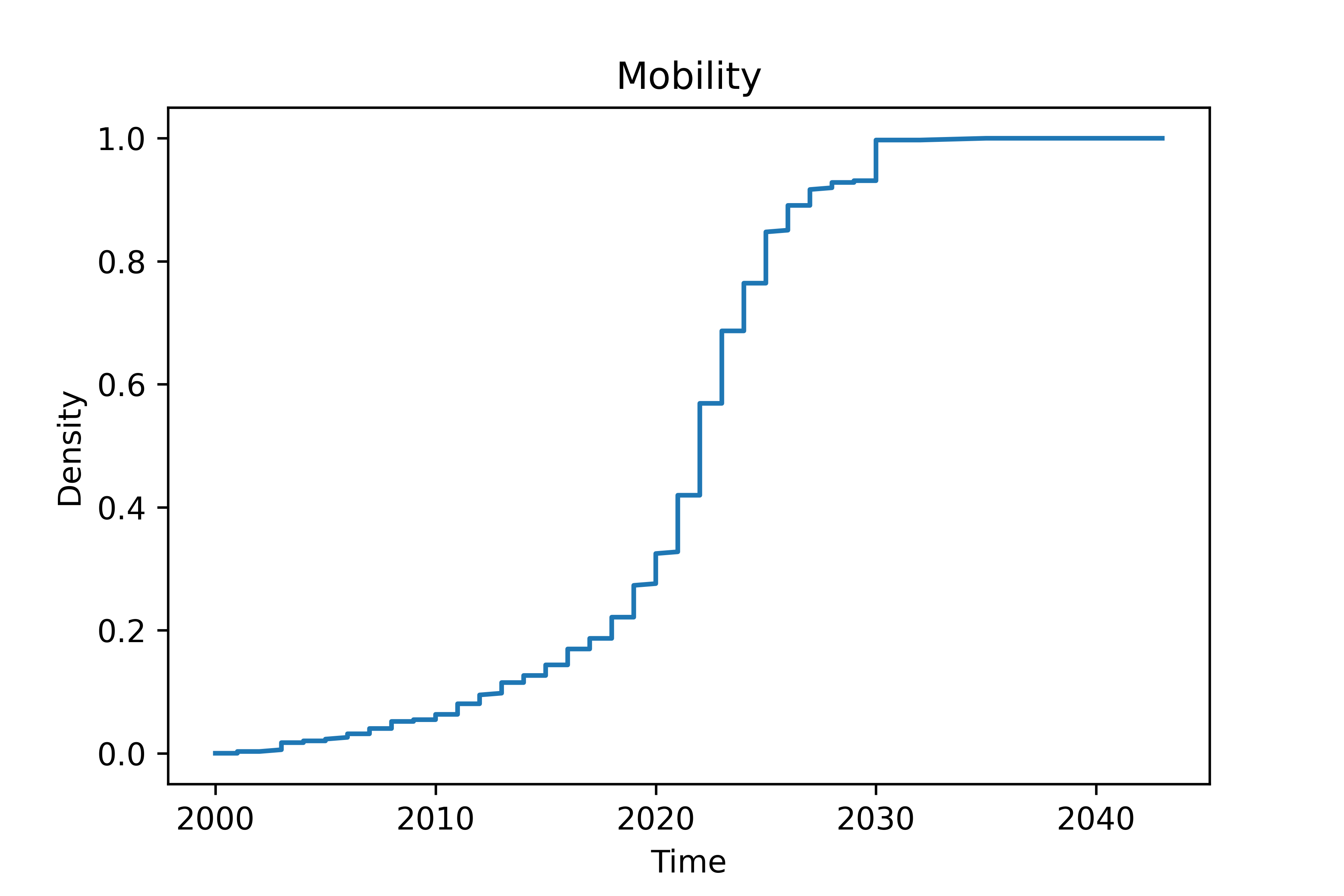}
        \caption{}
        \label{fig:CDF_Mobility}
    \end{subfigure}
\begin{subfigure}[b]{0.5\textwidth}
        \includegraphics[width=\textwidth]{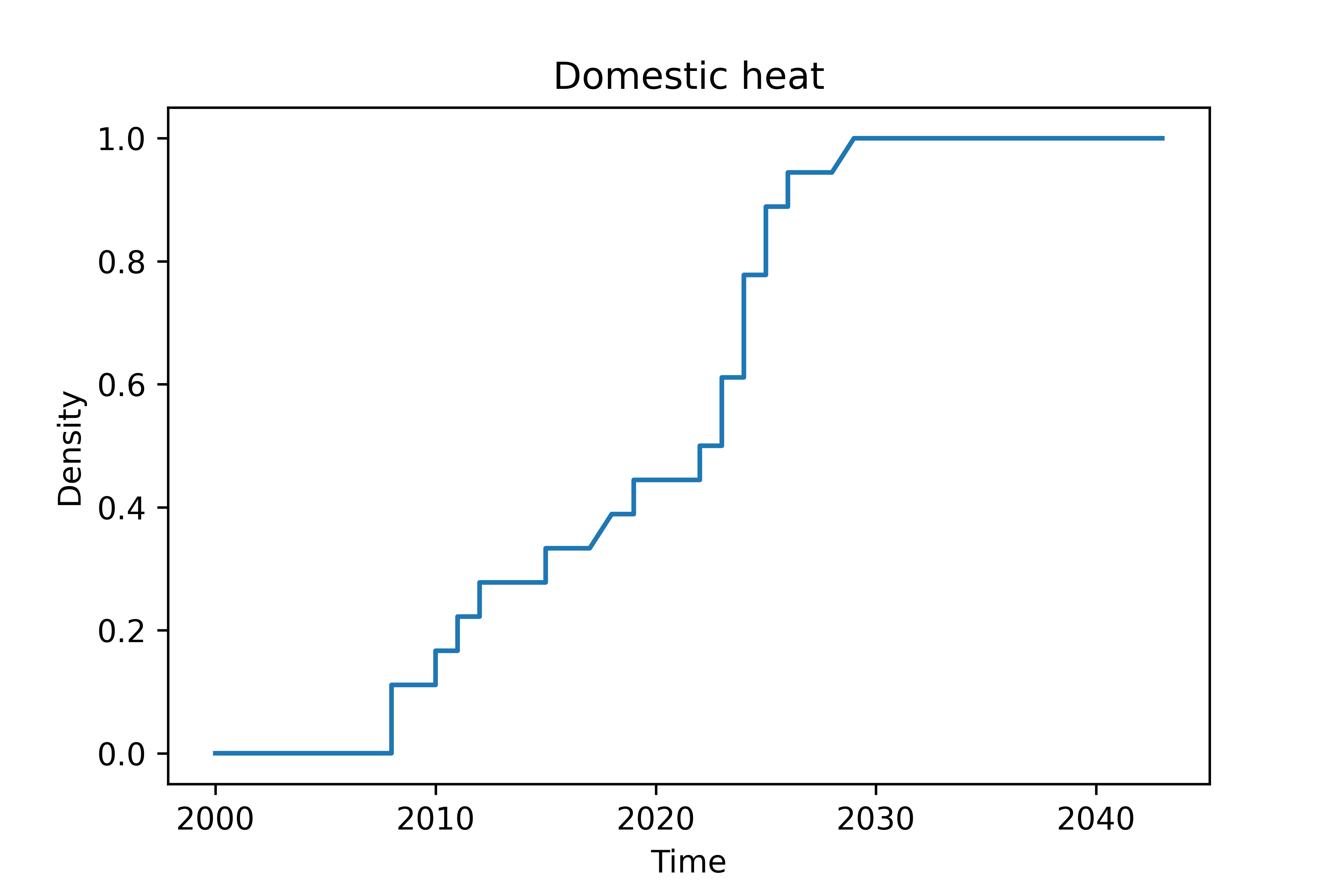}
        \caption{}
        \label{fig:CDF_Domesticheat}
    \end{subfigure}
\begin{subfigure}[b]{0.5\textwidth}
        \includegraphics[width=\textwidth]{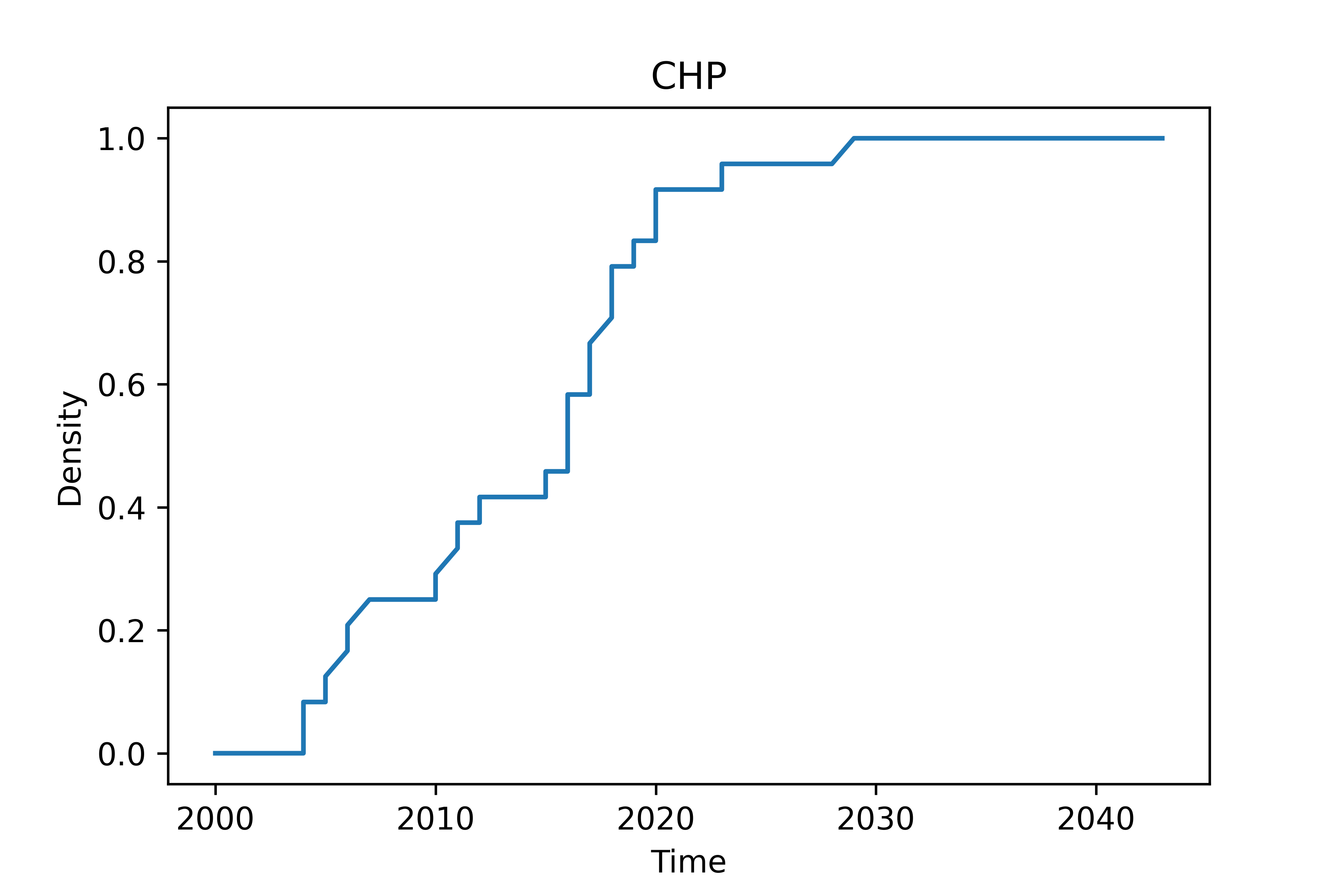}
        \caption{}
        \label{fig:CDF_CHP}
    \end{subfigure}
\begin{subfigure}[b]{0.5\textwidth}
        \includegraphics[width=\textwidth]{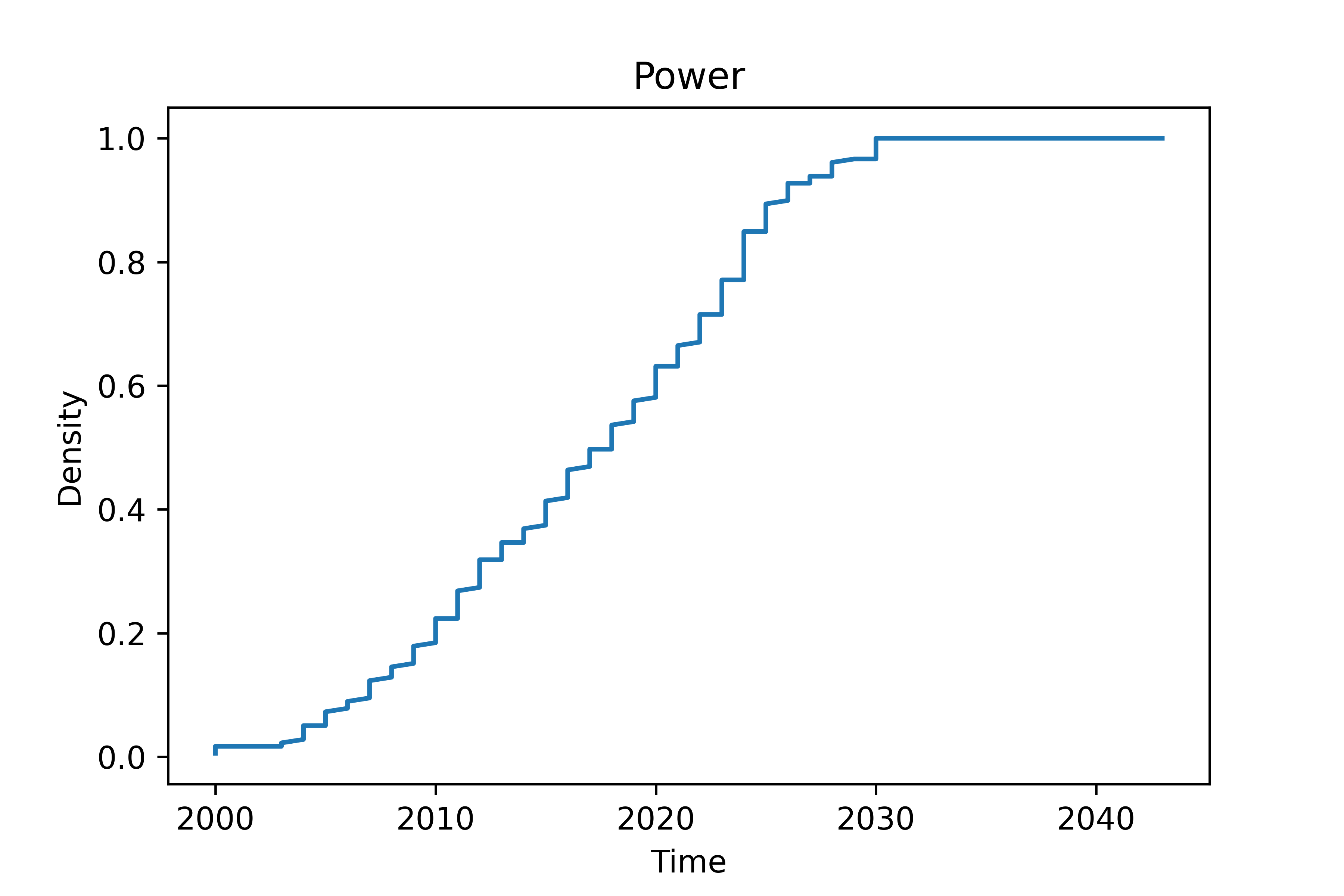}
        \caption{}
        \label{fig:CDF_Power}
    \end{subfigure}
    \caption{Cumulative distribution functions $F_S$ for eight sectors $S$, (\textbf{a}) refining (\textbf{b}) ammonia (\textbf{c}) synfuels (\textbf{d}) methanol (\textbf{e}) mobility (\textbf{f}) domestic heat (\textbf{g}) CHP (\textbf{h}) power. Sectors are described in Section~\ref{sec:data}. The~greatest collective similarity is observed between industrial applications, with~an explosion of planned plants in the 2020's. Power serves as an anomaly with its highly uniform trend of new~plants.}
    \label{fig:CDFs}
\end{figure}

In this section, we study the evolution of the propagation of low-carbon hydrogen plants for each sector and~investigate the similarities in these trends. Our primary mathematical object of study is the cumulative distribution function (CDF) $F_S(t)$ of number of plants over time for each end-use sector $S$. This is defined as the proportion of plants with end-use sector $S$ indexed with time $\leq t$ out of all plants with that sector. As~such, it is a non-decreasing function from 0 to 1. We remark that a small number of plants have more than one usage sector, and~thus will contribute to more than one CDF. This is no problem for our analysis. The~temporal CDFs for eight sectors $S$ are displayed in Figure~\ref{fig:CDFs}. As~only four plants in total are indexed before $t=2000$, we exclude these from our~figures.

\begin{figure}[H]
    
        \includegraphics[width=\textwidth]{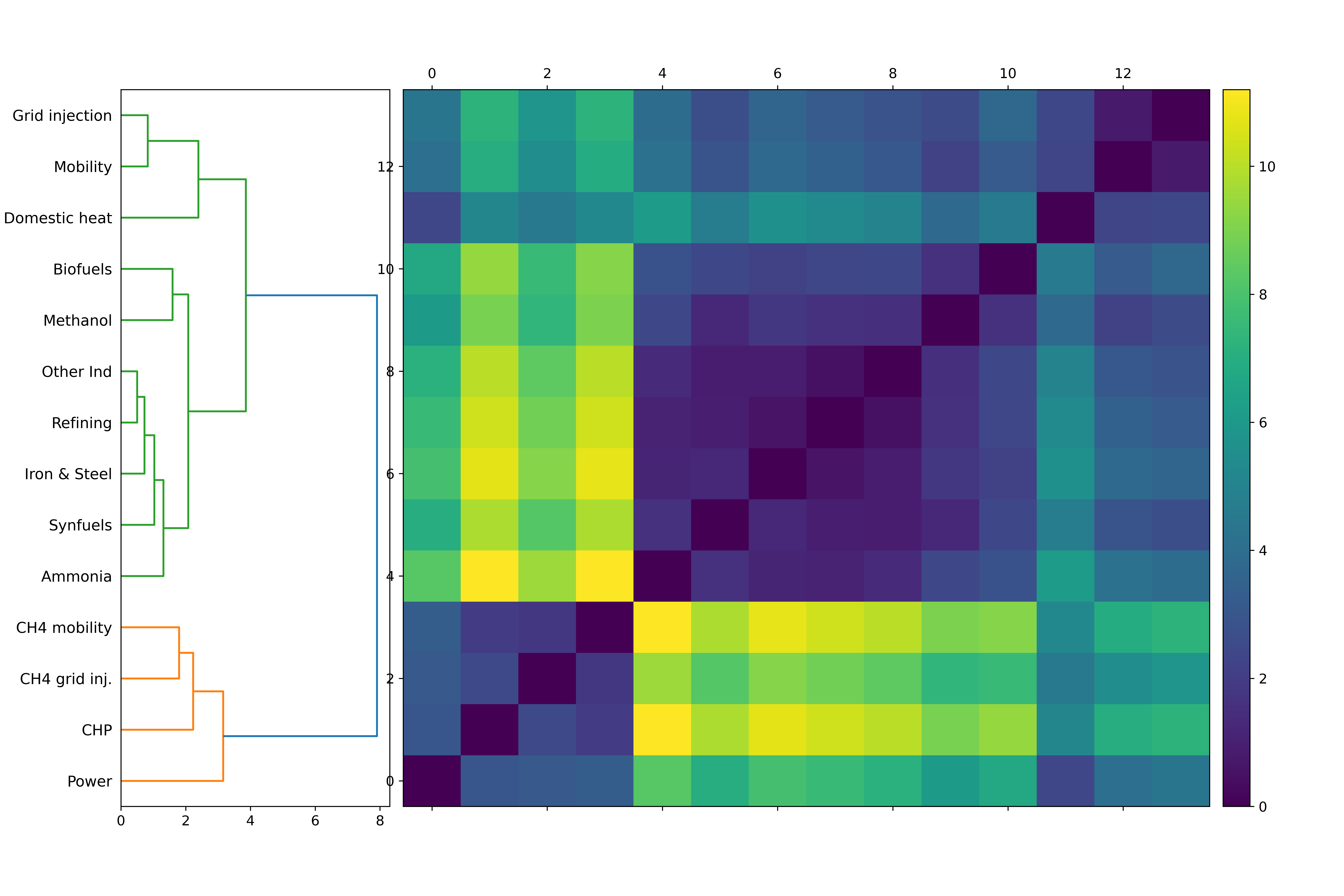}
    \caption{Hierarchical clustering on cumulative distribution functions $F_S$ (relative to time) for all 14 end-use sectors $S$ in our database. A~strong cluster of similarity is observed for the seven industrial uses, ranging from biofuels to ammonia. A~secondary cluster of more `consumer uses' is revealed from grid injection to domestic heat, whereas power is observed as an outlier due to its highly uniform~nature.}
   \label{fig:CDF_dend}
\end{figure}

In Figure~\ref{fig:CDF_dend}, we perform hierarchical clustering on these figures using the $L^1$ metric between these CDFs, defined as
\begin{align}
d(S_1,S_2)=\| F_{S_1} - F_{S_2}\| = \sum_{t} |F_{S_1}(t) - F_{S_2}(t)|.
\end{align}

As all CDFs take comparable values between 0 and 1, it is appropriate to directly compare them as~such.

Hierarchical clustering \cite{Ward1963, Szekely2005} is an iterative clustering technique that seeks to build a hierarchy of similarity between elements where there is some way to measure distance between them. In~this case, our elements are the cumulative distribution functions equipped with the distance above. This is opposed to a technique such as k-means, which ultimately specifies $k$ discrete groupings of elements. Hierarchical clustering is either agglomerative, where each element (CDF in our case) begins in its own cluster and branches between them are successively built, or~divisive, where all elements begin in one cluster and are successively split. The~results of hierarchical clustering are commonly displayed in dendrograms, which resemble branching trees. In~this paper, we implement agglomerative hierarchical clustering with the average linkage method~\cite{Mllner2013}. Figure~\ref{fig:CDF_dend} reveals numerous similarities and differences in the temporal distributions of new plants, which can also be seen in Figure~\ref{fig:CDFs}.

Examining Figure~\ref{fig:CDF_dend}, there is a dominant cluster of considerable similarity consisting of seven usage sectors: refining (Figure \ref{fig:CDFs}a), ammonia (Figure \ref{fig:CDFs}b), synfuels (Figure \ref{fig:CDFs}c), iron and steel, and other industry all occupy a subcluster of extreme similarity, while biofuels and methanol (Figure \ref{fig:CDFs}d) are deemed to be nearby. Curiously, all seven of these end-use sectors comprise industrial applications. Examining the plots in Figure~\ref{fig:CDFs}, we can see these sectors' temporal distributions are dominated by plants to be built in 2020--2030, with~a smaller number in the 2010's. Next, there is a moderately similar cluster consisting of mobility, grid injection and domestic heat. The~first two, as~represented by mobility (Figure \ref{fig:CDFs}e) have a very similar concave shape with a gradually increasing number of plants from 2000 to the 2020's, while domestic heat (Figure \ref{fig:CDFs}f) features slightly more activity in the 2010's. These three sectors reflect uses of hydrogen production that more directly service regular~consumers.

Next, there are some detected outliers. CHP (Figure \ref{fig:CDFs}g), CH4 mobility, and CH4 grid injection all share a similar shape, exhibiting a growth in plants earlier relative to all other sectors. The~most prominent outlier is power (Figure \ref{fig:CDFs}h), with~a highly uniform growth in new plants from 2000 to 2030 observed in no other sector. Indeed, power has been one of the most consistently popular uses of hydrogen production since its inception, and~this is reflected in the consistent and ongoing construction of new plants in this~sector.

\section{Usage distributions}
\label{sec:usagedistributions}
In this section, we investigate the distribution of plants across different sectors, focusing on each continent of location as well as green vs. blue/fossil technology. For~each plant in our dataset, the~location is recorded as either a country of location or continent. Thus, we first collate the plants on a continent-by-continent basis. We divide plants into seven continental and/or geopolitical regions as follows: Europe, North America (the United States and Canada), Latin America (including Mexico), Oceania, East Asia (China, Japan and Korea), Africa, and other (plants from elsewhere in Asia and the Middle East). We select these regions to combine countries according to both geographic proximity, political relations, and~economic development. Next, we divide every plant up according to groups that specify both a continental region and whether the technology is green or blue/fossil. For~example, we group all green European plants, or~all East Asian fossil plants. There are no African fossil or Latin American fossil plants in our dataset, so this leaves 12 different continental/technological groupings $G$.

For each group $G$ and end-use sector $S$, let $p^{(G)}_S$ be the proportion of plants devoted to end-use sector $S$ from that group, as~a fraction of all the plants in the group (counting a plant multiple times if it has multiple sectors). This forms a probability vector $\mathbf{p}^{(G)} \in \mathbb{R}^{14}$ of length 14. Some plants may have more than one use sector; this is fine. In those cases, a plant's multiple sectors are counted each time in the different coefficients of $p^{(G)}_S$ and also in the denominator. For~example, if~a group $R$ contained just one plant with two usage sectors, say refining and ammonia, then the vector $\mathbf{p}^{(R)}$ would be the vector $(\frac12, \frac12, 0,0,...,0) \in \mathbb{R}^{14}.$

Analogously, let $\mathbf{p}^{(H)}$ be the usage sector distribution vector for a different continental/technological group of plants $H$. Now we may define a distance between two distributions of usage sectors as h:
\begin{align}
\label{eq:distributiondist}
    d(G,H)= \|\mathbf{p}^{(G)} - \mathbf{p}^{(H)} \|_1 =\frac12 \sum_{S} |p^{(G)}_S - p^{(H)}_S|.
\end{align}
where the sum is taken over all 14 sectors $S$. This has the property that $d(G,H)=0$ if and only if groups $G$ and $H$ have an identical distribution of usages, again allowing multiple end-use sectors in some plants. Furthermore, $d(G,H) \leq 1$ is the maximal possible distance, with~equality if and only if the usage distributions are disjoint, with~no usage sectors in common at all between the two groups. When $G$ is empty, we set $\mathbf{p}^{(G)}$ to be the zero vector $\mathbf{0} \in \mathbb{R}^{14}$ which is  a distance of 1 from every other distribution and is thus always appearing as an outlier in hierarchical~clustering.

While seemingly simple, this distance may be interpreted as possibly the most suitable distance between two distributions of discrete sets, as~it can be shown to be equivalent to the discrete Wasserstein metric between distributions on a discrete metric space. More details are provided in~\cite{James2021_geodesicWasserstein} and Appendix \ref{app:discreteWass}.

\begin{figure}[H]
    
\begin{subfigure}[b]{\textwidth}
        \includegraphics[width=\textwidth]{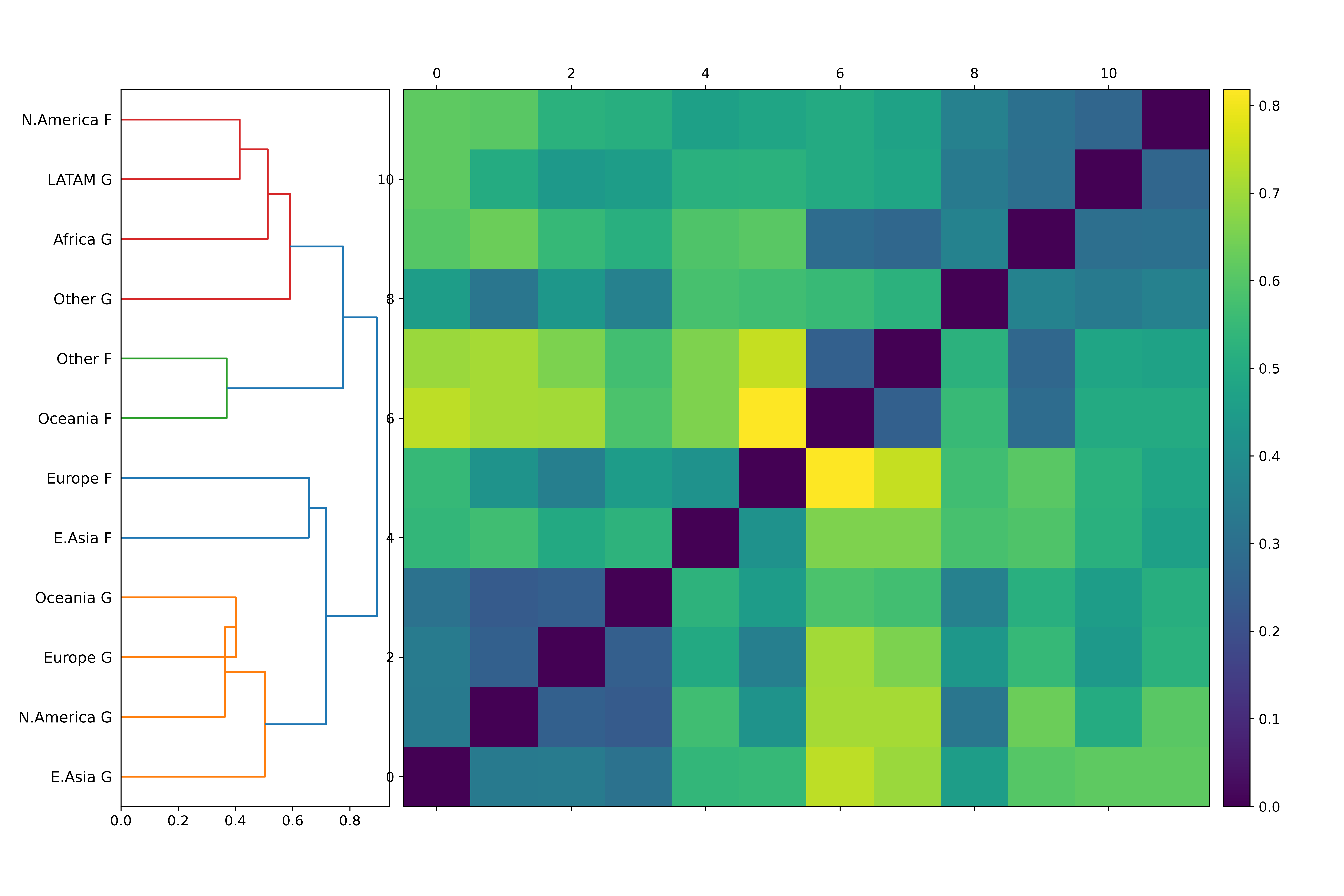}
        \caption{}
        \label{fig:distplotall}
    \end{subfigure}
\begin{subfigure}[b]{0.49\textwidth}
        \includegraphics[width=\textwidth]{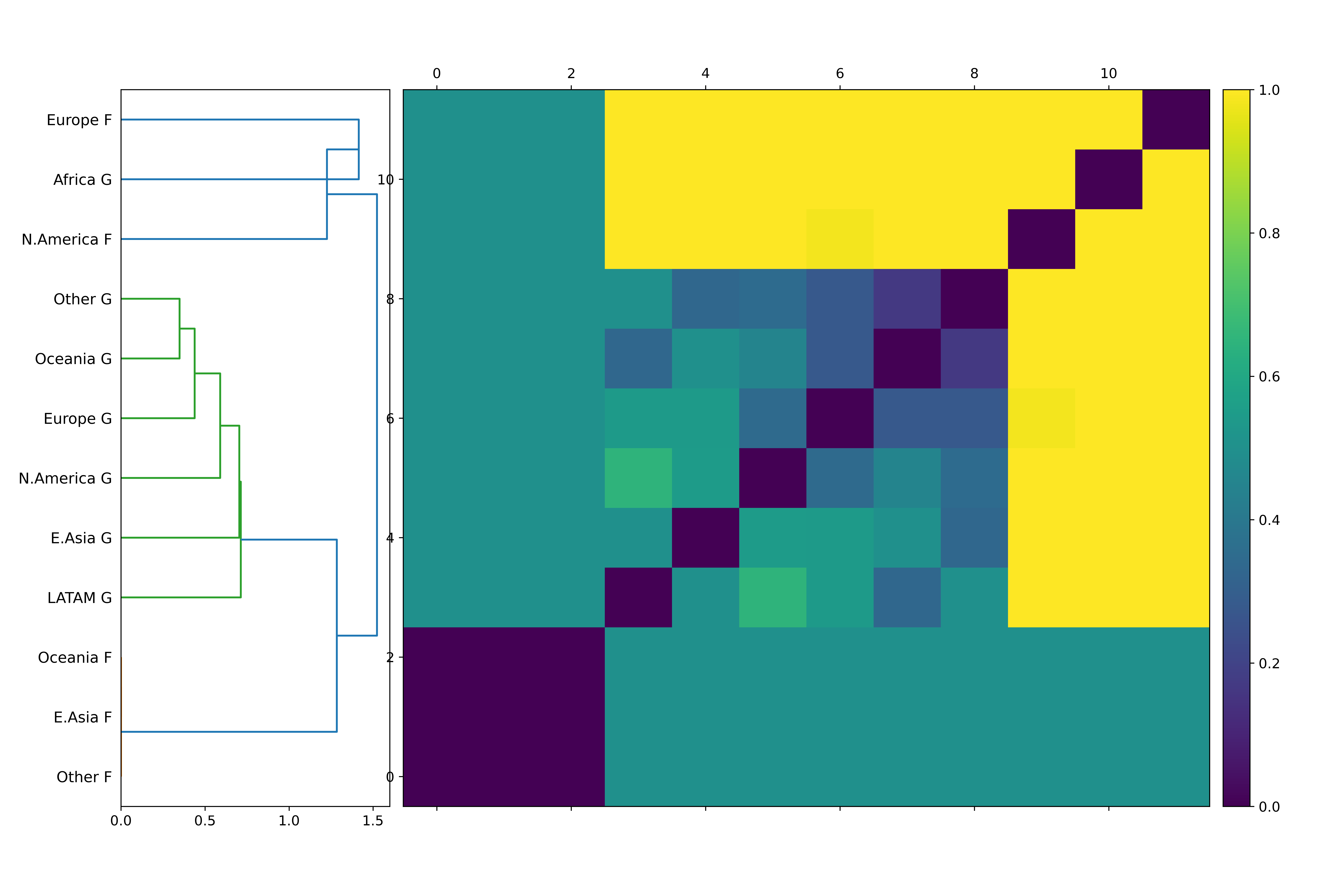}
        \caption{}
        \label{fig:distplot20002010}
    \end{subfigure}
    \begin{subfigure}[b]{0.49\textwidth}
        \includegraphics[width=\textwidth]{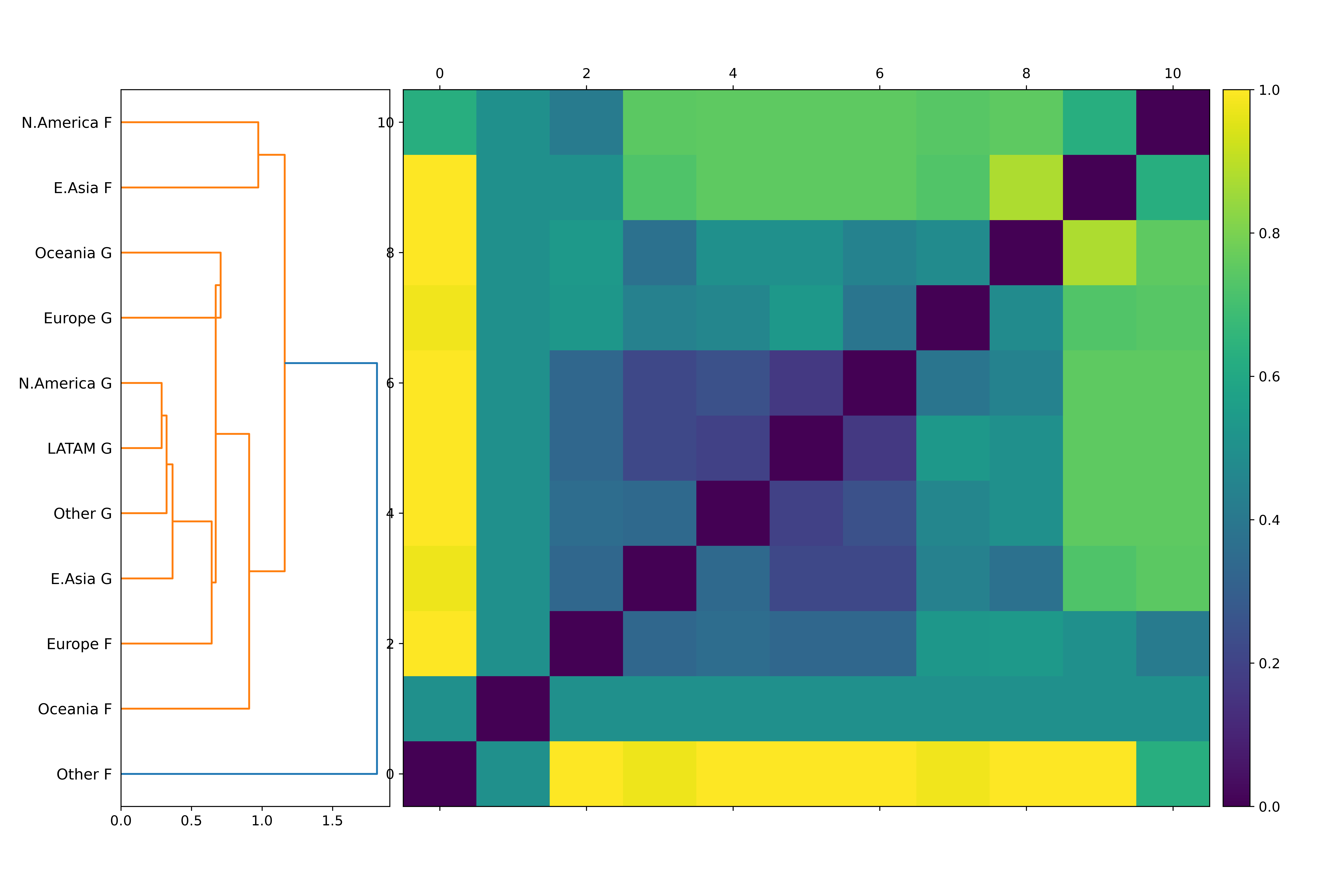}
        \caption{}
        \label{fig:distplot20102020}
    \end{subfigure}
        \begin{subfigure}[b]{0.49\textwidth}
        \includegraphics[width=\textwidth]{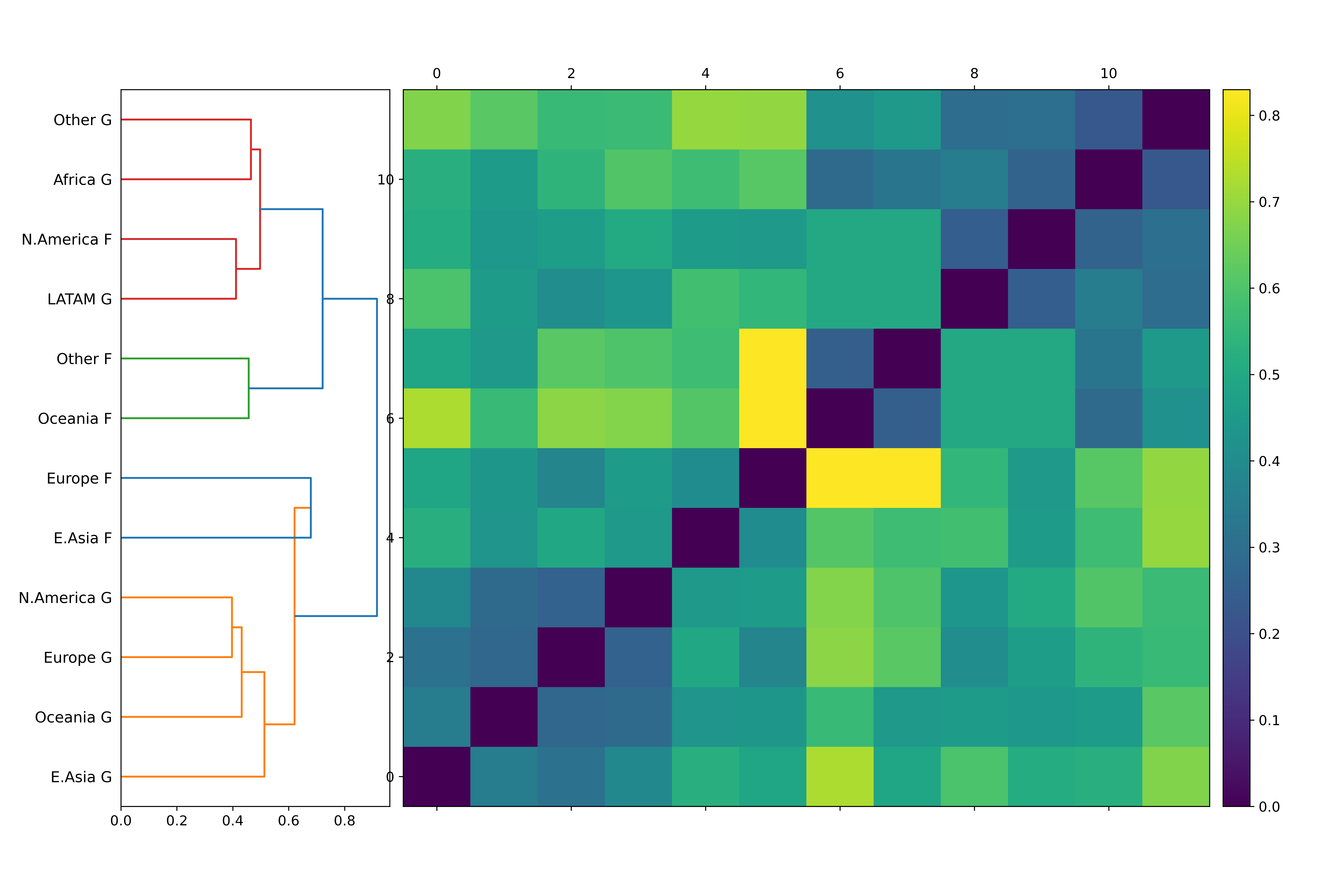}
        \caption{}
        \label{fig:distplot20202030}
    \end{subfigure}
\begin{subfigure}[b]{0.49\textwidth}
        \includegraphics[width=\textwidth]{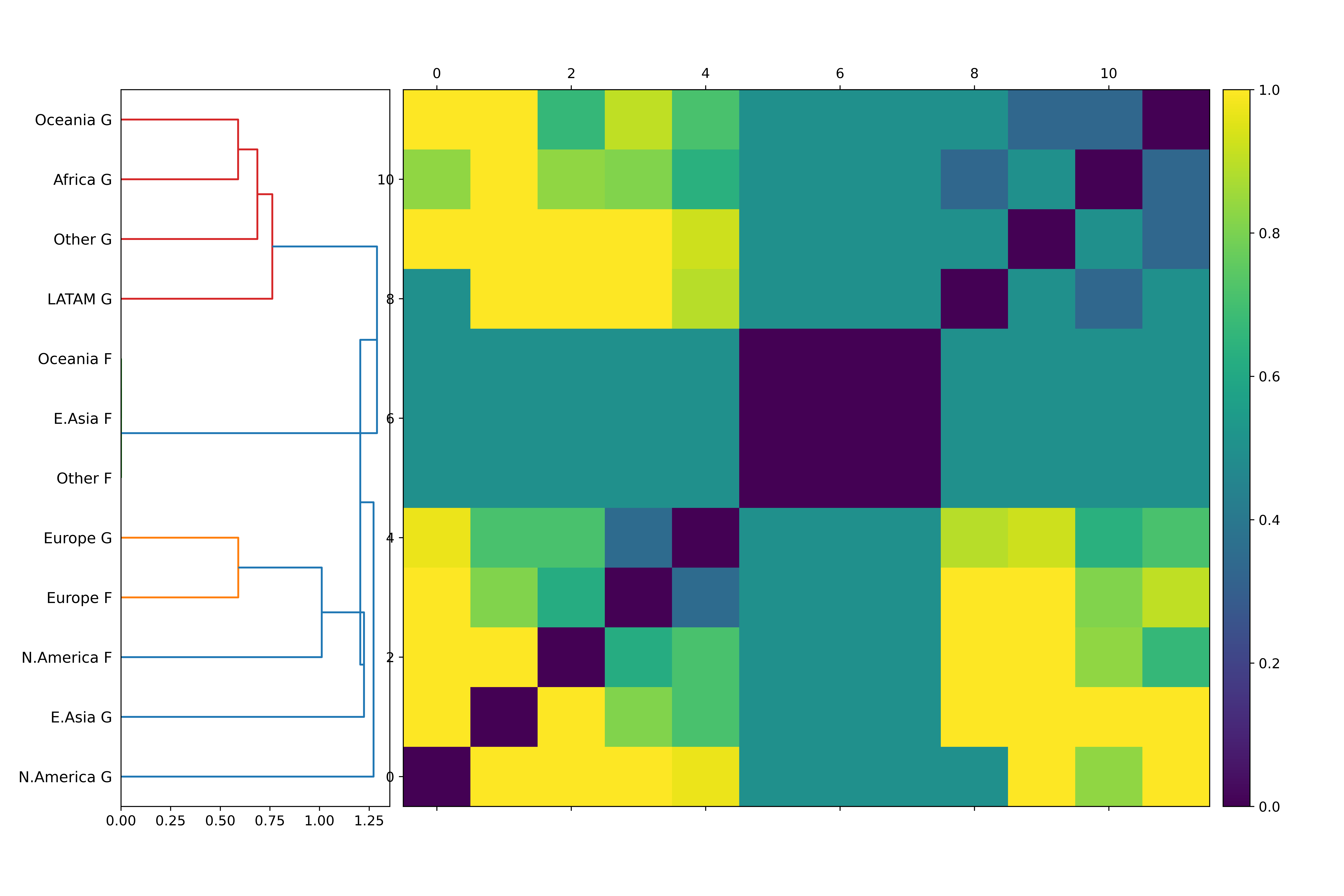}
        \caption{}
        \label{fig:distplot20302040}
    \end{subfigure}
    \caption{Distribution dendrograms between continental/technological groups $G$, produced by hierarchical clustering on the distance (\ref{eq:distributiondist}) for (\textbf{a}) the entire period of analysis (\textbf{b}) 2000---2009 \mbox{(\textbf{c}) 2010--2019} (\textbf{d}) 2020--2029 (\textbf{e}) 2030--2039. There are only 12 groups $G$ as there are no fossil plants in \mbox{Africa or Latin~America}.}
    \label{fig:distplots}
\end{figure}

\begin{figure}[H]
    
\begin{subfigure}[b]{\textwidth}
        \includegraphics[width=\textwidth]{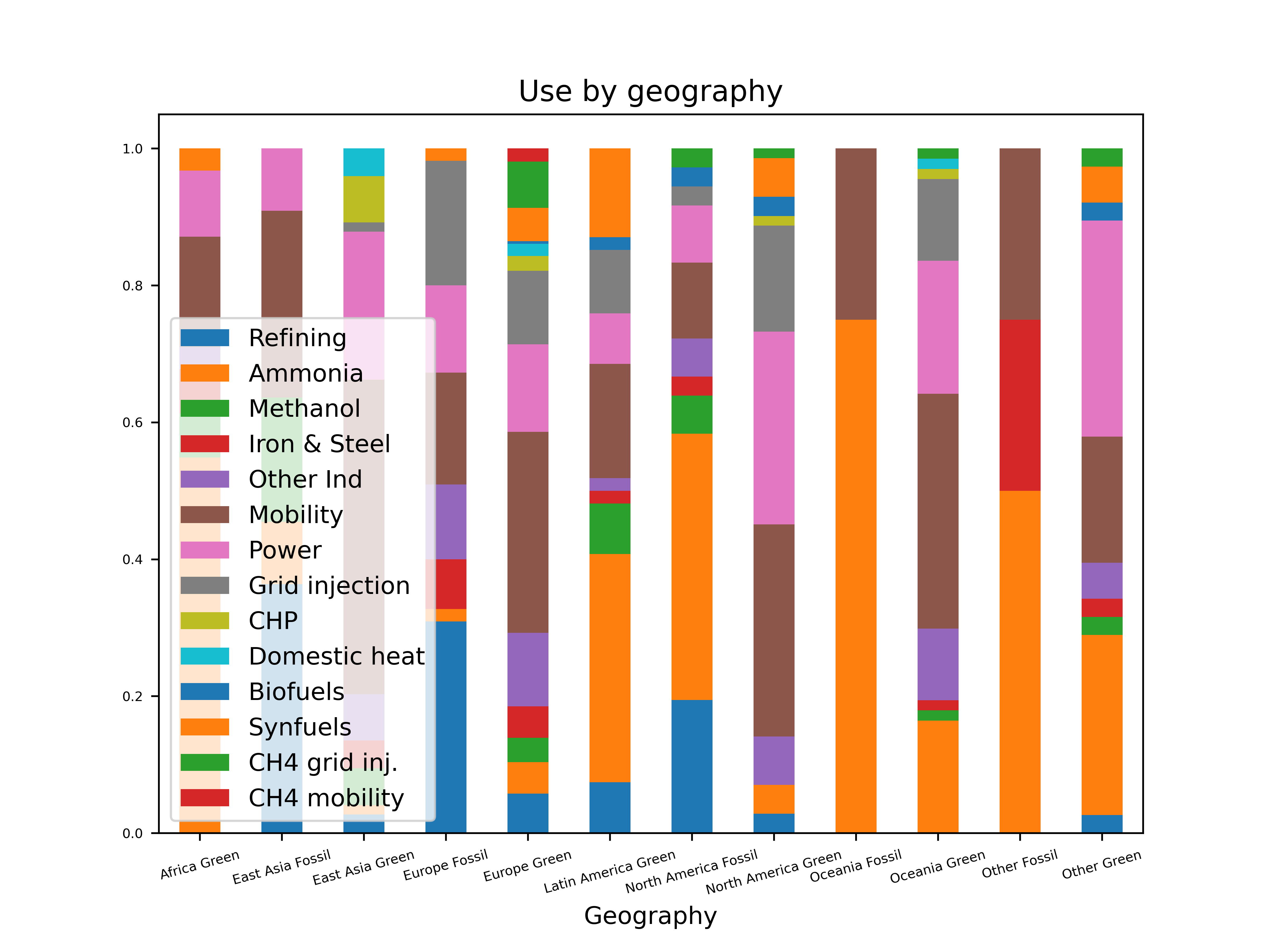}
        \caption{}
        \label{fig:barplotall}
    \end{subfigure}
\begin{subfigure}[b]{0.49\textwidth}
        \includegraphics[width=\textwidth]{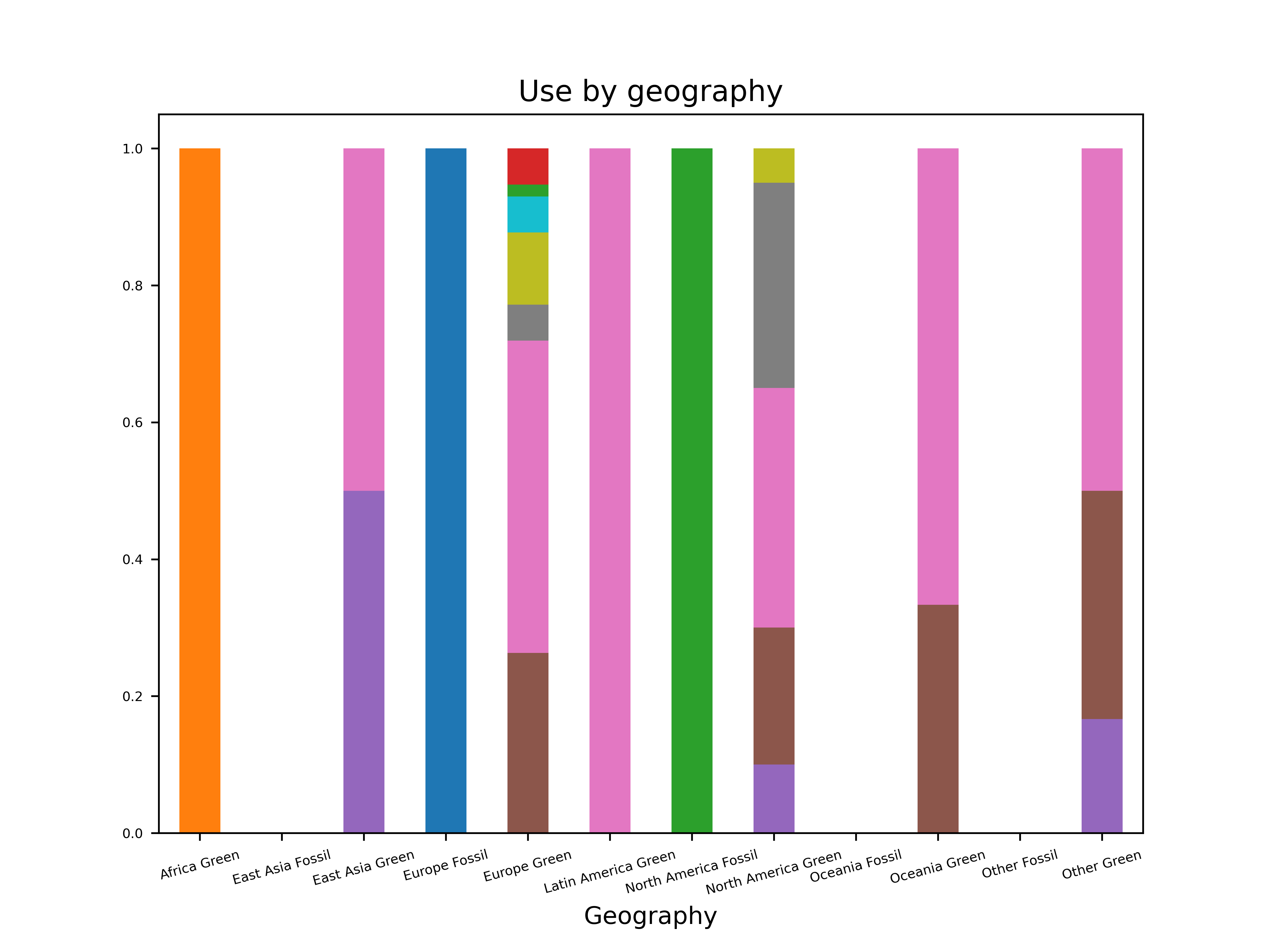}
        \caption{}
        \label{fig:barplot20002010}
    \end{subfigure}
    \begin{subfigure}[b]{0.49\textwidth}
        \includegraphics[width=\textwidth]{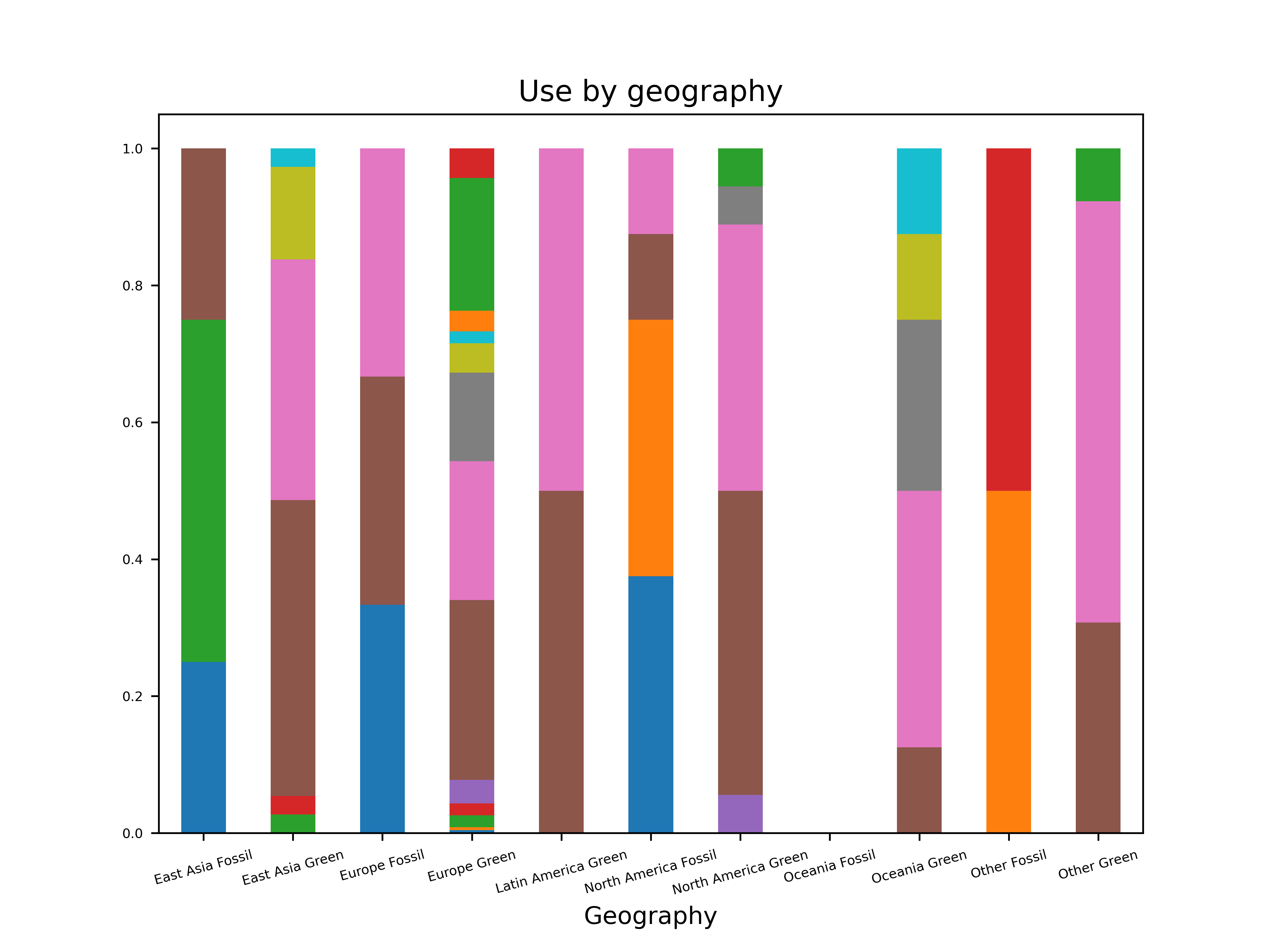}
        \caption{}
        \label{fig:barplot20102020}
    \end{subfigure}
        \begin{subfigure}[b]{0.49\textwidth}
        \includegraphics[width=\textwidth]{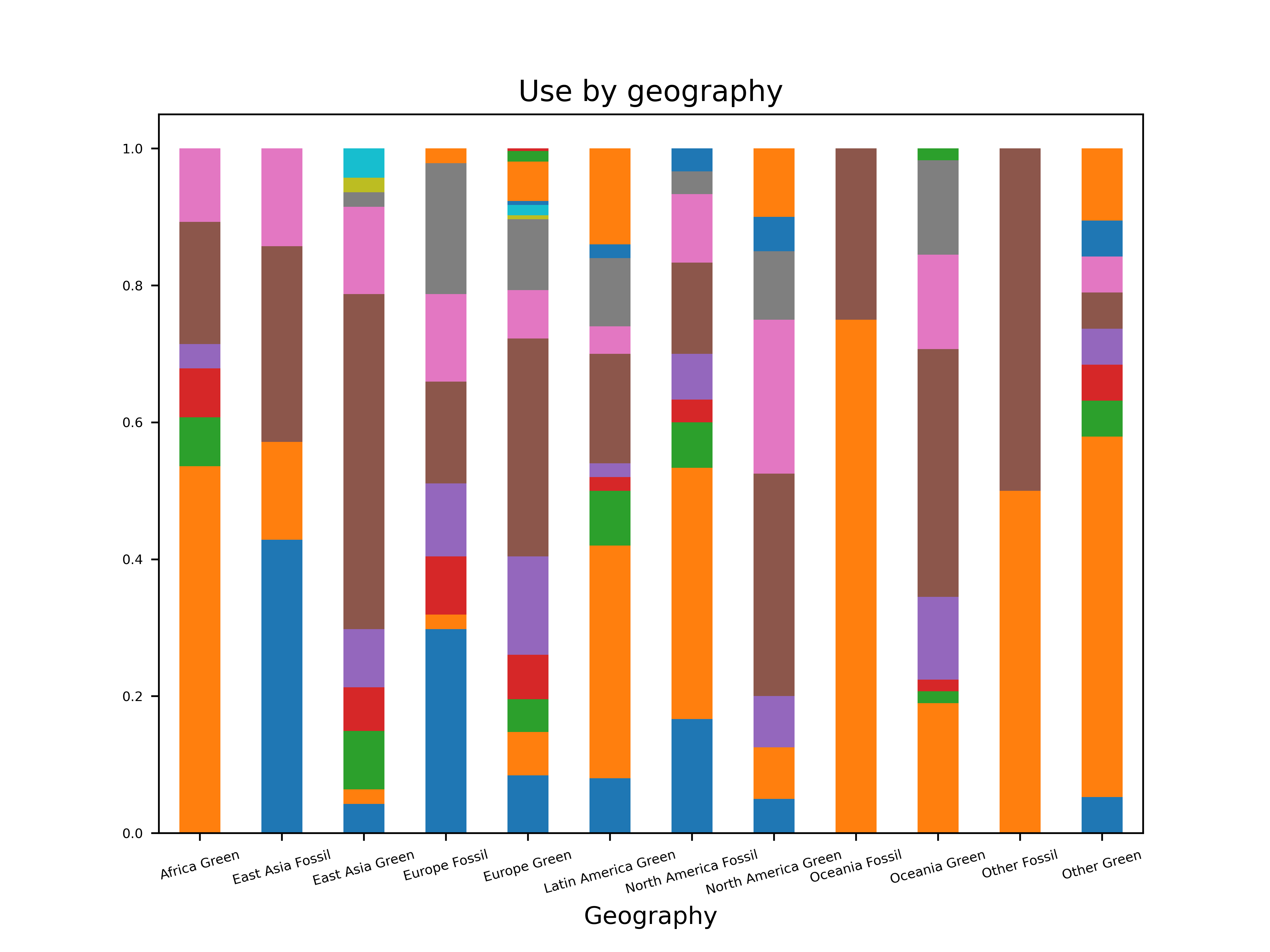}
        \caption{}
        \label{fig:barplot20202030}
    \end{subfigure}
\begin{subfigure}[b]{0.49\textwidth}
        \includegraphics[width=\textwidth]{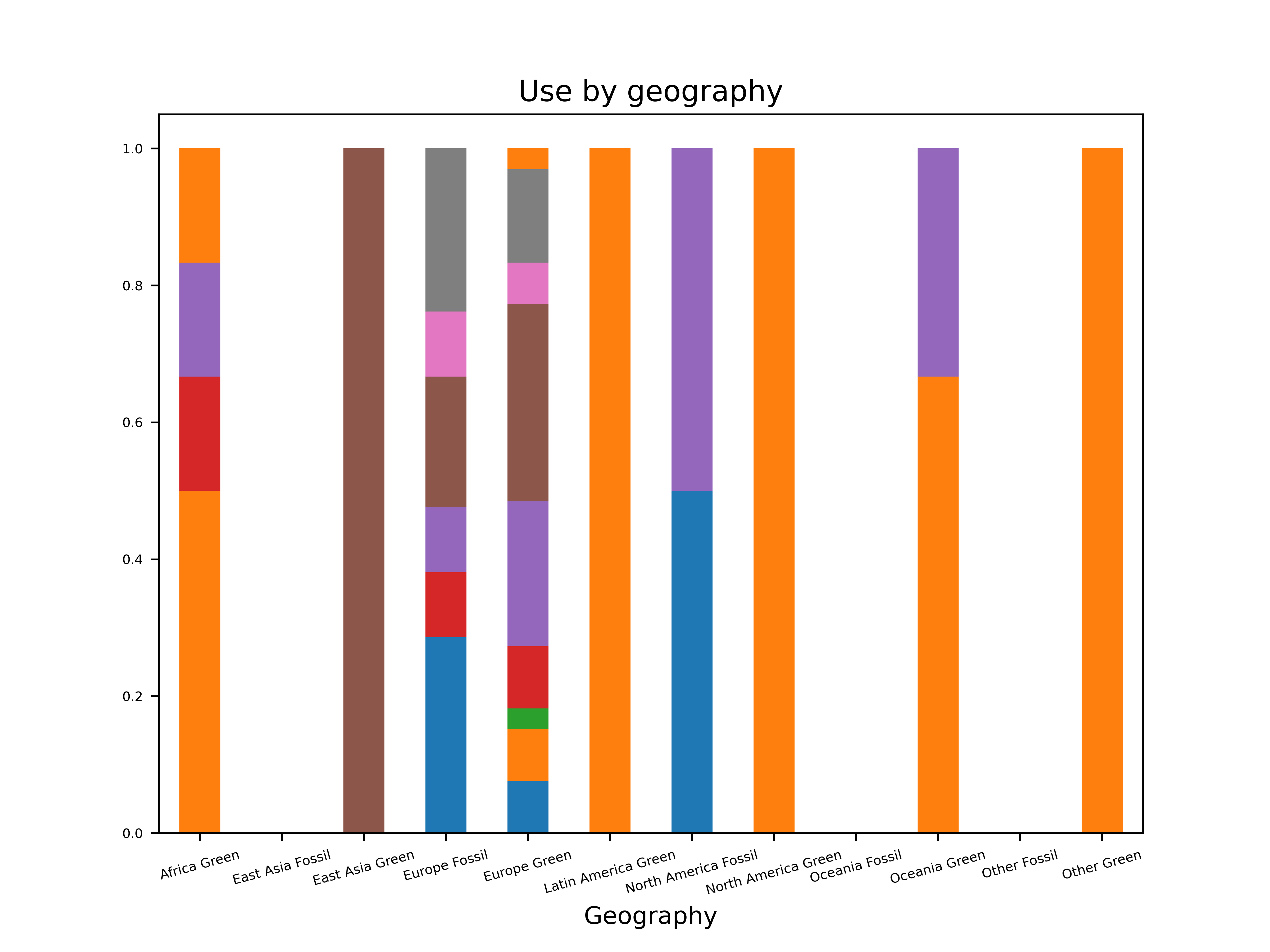}
        \caption{}
        \label{fig:barplot20302040}
    \end{subfigure}
    \caption{Stacked bar plots showing distribution of end-use sectors for (\textbf{a}) the entire period of analysis (\textbf{b}) 2000--2009 (\textbf{c}) 2010--2019 (\textbf{d}) 2020--2029 (\textbf{e}) 2030--2039. There are only 12 groups $G$ as there are no fossil plants in Africa or Latin America. Empty bars indicate no plants in that group over that~period.}
    \label{fig:barplots}
\end{figure}

In Figure~\ref{fig:distplots}, we perform hierarchical clustering on the 12 continental/technological groups $G$. Figure~\ref{fig:distplots}a uses data from the entire period of analysis, while Figure~\ref{fig:distplots}b--e only use data across a decade each, 2000--2009, 2010--2019, 2020--2029, and 2030--2039, respectively. We display stacked bar plots that graphically show the underlying distributions in Figure~\ref{fig:barplots}. Again, Figure~\ref{fig:barplots}a uses data from the entire period of analysis, and Figure~\ref{fig:barplots}b--e restrict their data to one decade~each.

Examining the dendrograms in conjunction with the stacked bar plots highlights the key similarities and differences that give rise to the structure of the clustering observed. In~Figure~\ref{fig:distplots}a, the~12 continental/technological groups divide into two primary subclusters of six groups each, with~no outliers observed. A~close subcluster of similarity is observed between Oceanian green, European green, North American green, and East Asian green plans. That is, we observe notable similarity in the usage distributions of green hydrogen plants from the four most economically developed continental groups. Examining Figure~\ref{fig:barplots}a, we see that these four groups are characterised by a distribution of end-use sectors that crosses (almost) all end-use sectors in each group. Furthermore, these four groups all have low proportion of plants dedicated to ammonia production, an~observation we will return to below. Next, European fossil and East Asian fossil plants are deemed similar and in the same primary cluster. Examining Figure~\ref{fig:barplots}a, we observe very low ammonia usage, but~not as diverse of a range of end-use sectors as the previous~subcluster.

Turning to the other main cluster, we see that all six of these groups, North American fossil, Latin American green, African green, other green, other fossil, and Oceanian fossil plants, have high proportions of plants dedicated to ammonia production. The~observed subcluster of other fossil and Oceanian fossil plants exhibits more plants dedicated to ammonia than any other group, while the remaining subcluster is slightly more evenly distributed. Thus, the~primary division of the 12 groups into two main clusters is almost entirely explained by a dramatic difference in just a single sector of usage: ammonia~production.

There are numerous insights to be gleaned from examining the dendrograms and bar plots on a decade-by-decade basis. In~the decade 2000--2009, a~subcluster of similarity is observed between six of seven continents' green power plants: Latin American green, East Asian green, other green, Oceanian green, European green and North American \mbox{green (Figure \ref{fig:distplots}b)}. Turning to Figure~\ref{fig:barplots}b, we can see that this similarity can be primarily explained by the prevalence of power generation plants among these green plants. In~the next \mbox{decade (Figure \ref{fig:distplots}c)}, the~subcluster structure of North American green, Latin American green, other green, and East Asian green plants (with European fossil, Oceanian green, and European green plants being slightly less similar) is primarily explained with a high proportion of mobility and power among these plants. The~decade 2020--2029's clustering (Figure \ref{fig:distplots}d) is nearly identical to that of the full period, as~the counts of the full period are dominated by this decade, so the distributional differences between groups of 2020--2029 closely resemble those over all time. Finally, the~decade 2030--2039 is mostly dominated by missing data or continental/technological groups that have only planned to build plants dedicated to ammonia production. As~this is quite far into the future, it will of course be subject to change as more plants and their end-use sector are planned and~built.

\section{Trends in capacity over time and relative to technology and end-use sector}
\label{sec:regression}

\begin{figure}[H]
    
        \includegraphics[width=\textwidth]{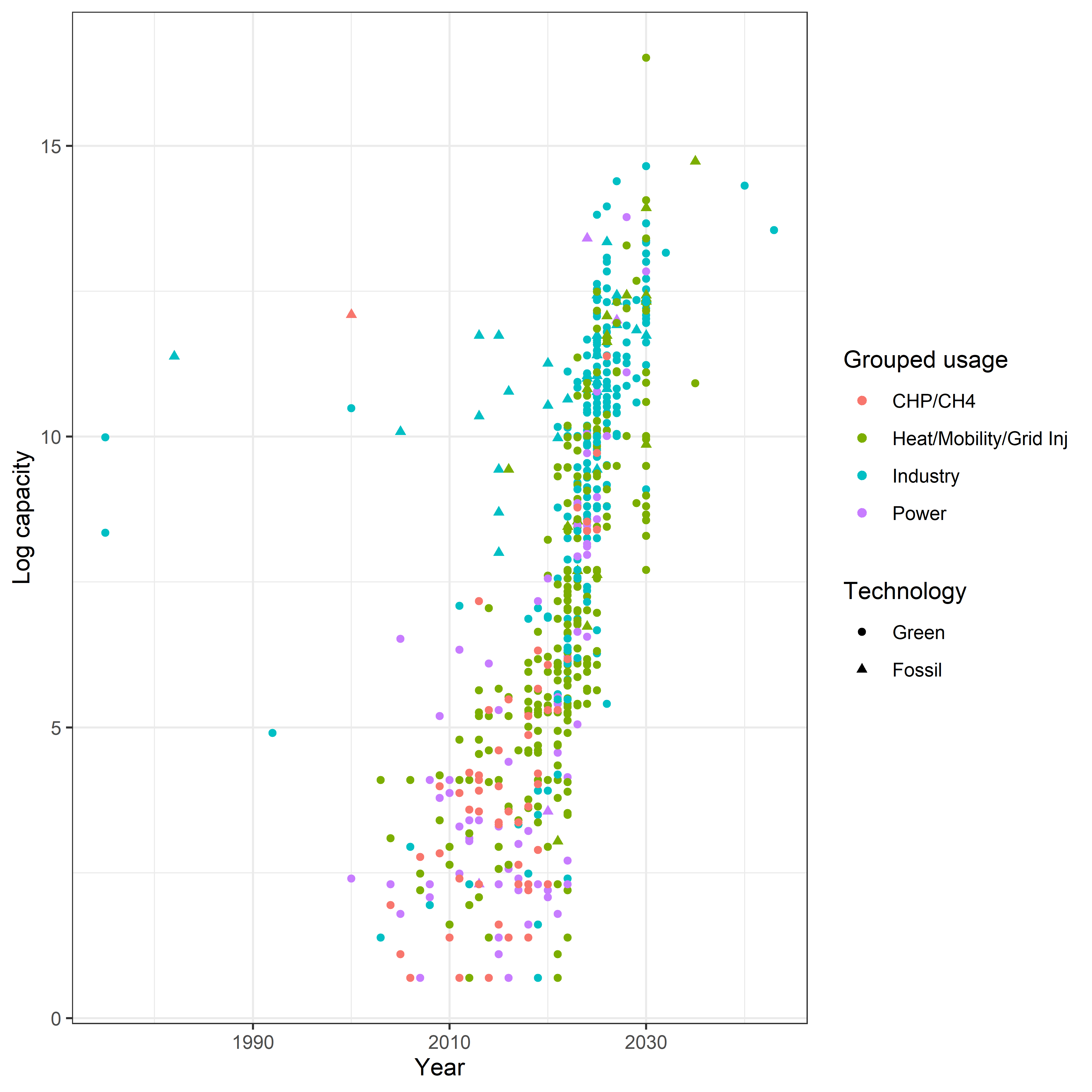}
    \caption{Log capacity vs. year of construction for all plants in our dataset with available data. We classify plants by both their technology as well as their usage, using the clusters of end-use sectors from Section~\ref{sec:CDFs} Figure~\ref{fig:CDF_dend}. We can see the early dominance of blue plants (with a few early exceptions) by several orders of magnitude, but~this is closing with~time.}
   \label{fig:capacity_scatter}
\end{figure}

In this section, we examine the increase in capacity of low-carbon hydrogen plants over time while examining differences between technology and end-use sector. In~Figure~\ref{fig:capacity_scatter}, we display all plants with an identified technology, end-use sector, and capacity in our dataset. Displaying the logarithm of the capacity against the year (or projected year) of construction, we see an approximate linear trend between the log of capacity and time, suggesting an exponential increase in capacity over time. We separate plants' hydrogen technology into green and blue/fossil and~separate them into end-use sectors. For~graphical readability, we have chosen to separate sectors into the four clusters discussed in Section~\ref{sec:CDFs} that were determined in Figure~\ref{fig:CDF_dend}. These are: the seven industrial applications, the~three consumer/domestic sectors (domestic heat, grid injection, mobility), the~CHP/CH4 outlier cluster, and~the single outlier (power).

For robustness, we also repeated this plot with some slightly different groupings. For~example, we separated CHP from the two CH4 uses, and~we separated domestic heat from mobility/grid injection (as domestic heat is not quite as similar as the latter two in \mbox{Figure~\ref{fig:CDF_dend}).} No substantial differences were observed in the log capacity vs. year scatter plot, so we proceed with the groupings from Figure~\ref{fig:CDF_dend} as they~are.

To confirm this exponential fit, we implemented a series of linear regressions between the (log) capacity, time, as~well as technology and usage. {Linear regression is a commonly used and celebrated statistical model. This method seeks to model a response or target variable (in this case capacity or log capacity) as a linear function of other variables, including time. The~most suitable linear coefficients are determined by a process called least squared estimation~\cite{ISL}, where the line of best fit is defined by the property that the sum of square deviations from that line is minimised. Linear regression is perhaps the simplest of all statistical or machine learning models; however, it is by no means trivial. In~its simplicity, it can provide greater interpretability than much more complex models, and~sometimes perform just as well. That is our aim today. We remark that our purpose is not predictive (predicting future trends in hydrogen plants) but to describe trends observed in the data as of now (including planned plants for future construction). Our linear regression models} take one of the following forms:
\begin{align}
\label{eq:linearlinear}
   y_i &= \beta_0 + \beta_1 t_i + \sum_{j} \beta_j^{use} \bm{1}_{ij}^{use} + \beta^{tech} \bm{1}_{i}^{tech},\\
   \label{eq:loglinear}
       \log( y_i) &= \beta_0 + \beta_1 t_i + \sum_{j} \beta_j^{use} \bm{1}_{ij}^{use} + \beta^{tech} \bm{1}_{i}^{tech}, 
\end{align}
where $y_i$ is the capacity, $t_i$ is the year for each plant, and~we include dummy variables for end-use sectors and technology (green or fossil). With~only two technologies, only one dummy variable is necessary; for the end-use sectors, we require one fewer dummy variable than the number of sectors. We run eight linear regressions across three choices---capacity or log capacity, all 14 end-use sectors, or the four grouped usages as in \mbox{Figures~\ref{fig:CDF_dend} and \ref{fig:capacity_scatter}}---and the inclusion of the technology variable or not. We record the adjusted $R^2$, a~measure of goodness of fit, for~all eight regressions in Table~\ref{tab:adjustedR2}.

\begin{table}[H]
\caption{Adjusted $R^2$, a~measure of goodness of fit for~eight different linear regression models between (log) capacity, grouped or ungrouped sectors, and~technology. A~far better fit is observed for the exponential~models.}
\label{tab:adjustedR2}
\setlength{\tabcolsep}{2.9mm}{ 
\begin{tabular}{ccccc}
\toprule
 \multirow{2}{*}{\textbf{Adjusted} \boldmath{$R^2$}} & \multicolumn{2}{c}{\textbf{No Separation by Tech}} & \multicolumn{2}{c}{\textbf{Stratified by Tech}} \\
 & \textbf{Grouped Sectors} &\textbf{ All Sectors} & \textbf{Grouped Sectors} & \textbf{All Sectors} \\
  \midrule
Capacity & 0.0255 & 0.0204 & 0.0273 & 0.0226 \\ 
Log capacity & 0.569 & 0.587 & 0.607 & 0.622 \\ 
  \bottomrule
\end{tabular}}

\end{table}

The results confirm the main finding visible in Figure~\ref{fig:capacity_scatter}, a~clearly superior fit  exists when considering the log capacity model across all other choices. This strongly suggests a rather regular exponential growth in the capacity of plants over time, including when separating across usage sectors. {That is, we have used goodness-of-fit measures to reveal that the increase in capacity over time is better represented as exponential than linear, and~rather well represented as exponential in its own right.} Viewing Figure~\ref{fig:capacity_scatter} across the usage groups tells a promising story. With~the exception of the industrial sectors, which saw a relatively early onset of high-capacity plants (though few in number), the~other groups are growing at a commensurate rate, as~can be seen by the relative collinearity of domestic heat/mobility/grid injection, CHP/CH4, and~power. We can also see the welcome closing of the gap between green and blue/fossil hydrogen plants'~capacities.

\section{Conclusion}
\label{sec:conclusion}

This paper studies the capacity and end-use sectors of low-carbon hydrogen technologies over time, between green (zero-carbon) or blue/fossil \mbox{(captured and stored carbon)} technologies, and~differences across the world. Combining the findings from investigating the distinct phenomena explored in this paper provides a unique, holistic overview regarding the state of low-carbon hydrogen projects globally. We are unaware of any similar work in the application field of low-carbon~hydrogen.

In Section~\ref{sec:CDFs}, we investigate the distributions of new hydrogen plants over time. Our results show a welcome growth in new and planned plants in recent years and the 2020's, consistent across sectors. By~analysing the discrepancy between temporal cumulative distribution functions, we reveal interesting structural similarity, including a strong cluster of industrial applications, where the number of plants is presently increasing significantly, and~a smaller cluster of domestic applications. Some outlier end-use sectors are also revealed, particularly power, which has been much more consistent and uniform in its increase in new plants over time, heralding a more reliable and proven usage sector. Simply put, we classify end-use sectors as under uniform growth (power), concave-up growth (domestic applications), and~explosive growth (industrial applications).

In Section~\ref{sec:usagedistributions}, we study usage distributions of end-use sector across continental and technological groupings of hydrogen plants, both across the entire period of analysis and on a decade-by-decade basis. We reveal additional structural similarity, including the greatest diversity of usages among green power plants of the four most economically advanced continents, a~welcome sign of the widespread utility and diverse applications of hydrogen energy. Interestingly, single key end-use sectors can almost entirely explain the broad-level cluster structure, such as ammonia across the entire period and the 2020--2029 decade, power in the 2000--2009 decade, and~mobility and power in the 2010--2019~decade.

Finally, in~Section~\ref{sec:regression}, we demonstrate the changing relationship between green and blue hydrogen project capacity across a range of sectors (and the groupings determined by Section~\ref{sec:CDFs}). We observe both graphically and analytically an exponential growth in the capacity of green hydrogen projects across all sectors, both grouped and ungrouped. This is an excellent sign for the future of this renewable fuel and suggests that investing in green hydrogen projects will likely provide returns in the coming years, regardless of end-use sector. Indeed, the~diversity of the end-use applications, especially with exponentially increasing capacity of plants, means that the market for hydrogen technology is both growing and inherently diversified. This reflects well on the potential returns in governments and private entities investing in this technology, as~well as less risk than a technology with more limited~uses.

The future development of hydrogen plants must consider safety, profitability, and widespread utility of hydrogen production, all to further the key aim of decarbonisation, while blue hydrogen plants discussed in this manuscript attempt to sequester their carbon emissions, the~use of oil, coal, or natural gas in the process carries a significant risk of harmful byproducts, and~the capture of emissions is imperfect~\cite{IPCC}. Thus, we hope to see growth in the number of green hydrogen plants across all end-use sectors, providing useful benefits to a diverse range of societal sectors, and~associated increases in their capacity. Monitoring these trends may be crucial to the future of decarbonisation for many industries as diverse societies drive towards a sustainable~existence.

\authorcontributions{Both authors contributed equally in every aspect of the~paper. All authors have read and agreed to the published version of the manuscript.}

\funding{This research received no external funding.}

\dataavailability{The data analysed in this article are publicly available at~\cite{Hydrogendata}. {A cached copy is available at \url{https://github.com/MaxMenzies/HydrogenData}  (accessed on 26 November 2022).}}

\conflictsofinterest{The authors declare no conflict of~interest.}

\appendixtitles{yes} 
\appendixstart
\appendix
\section[\appendixname~\thesection]{Probability distribution distance}
\label{app:discreteWass}

In this appendix, we give a technical explanation of why our distance (Equation (\ref{eq:distributiondist})) can be interpreted as perhaps the most appropriate distance between two distributions on a discrete set. Let $(X,d)$ be a metric space, $\mu,\nu$ two probability measures on $X$, and~$q \geq 1$ a real number. The~Wasserstein metric between distributions $\mu, \nu$ is defined as
\begin{align}
\label{eq:Wasserstein}
    W^q (\mu,\nu) = \inf_{\gamma} \bigg( \int_{X \times X} d(x,y)^q d\gamma  \bigg)^{\frac{1}{q}},
\end{align}
where the infimum is taken over all probability measures $\gamma$ on $X \times X$ with marginal distributions $\mu$ and $\nu$, respectively. From~here, let $q=1$. By~the Kantorovich--Rubinstein formula~\cite{Kantorovich}, there is an alternative formulation  of Equation (\ref{eq:Wasserstein}) when X is compact\mbox{ (for example, finite):}
\begin{align}
\label{eq:Wassersteinalt}
    W^1 (\mu,\nu) = \sup_{F} \left| \int_{X} F d\mu - \int_{X} F d\nu  \right|,
\end{align}
\textls[-48]{where the supremum is taken over all $1$-Lipschitz functions $F: X \to \mathbb{R}$, meaning \mbox{$|F(x,y)| \leq d(x,y)$} for all $x,y$.}

\begin{Proposition}
Let $(X,d)$ be a finite set with the discrete metric, with~$d(x,y)=1$ for $x\neq y$ and 0 otherwise. Let $W^1(\mu,\nu)$ be the $L^1$-Wasserstein metric between two probability measures $\mu,\nu$ on $X$, as~expressed in Equation (\ref{eq:Wassersteinalt}). That is,
\begin{align}
\label{eq:Wassersteinalt_new}
    W^1 (\mu,\nu) = \sup_{F} \left| \int_{X} F d\mu - \int_{X} F d\nu  \right|.
\end{align}

\textls[-35]{Associate to $\mu$ and $\nu$ corresponding distribution functions or probability vectors $f$ \mbox{and $g$, respectively.}}

Then, the~supremum written above is optimised by the following choice of $F$:
\begin{align}
\label{eq:bestF_new}
F(x) = \begin{cases}
1, & f(x) \geq g(x), \\
0, & f(x)<g(x).
\end{cases}
\end{align}
Thus, $W^1(f,g)$ reduces to the same form of Equation (\ref{eq:distributiondist}), namely
\begin{align}
\label{eq:discreteWass_new}
W^1(f,g)=\frac12 ||f - g||_1.
\end{align}
\end{Proposition}
\begin{proof}
Let $F$ be an arbitrary $1$-Lipschitz function on $X$ with its discrete metric. Thus, $F: X \to \mathbb{R}$ and $|F(x) - F(y)| \leq d(x,y) \leq 1$ for all $x,y$. We define $M$ and $m$ by\mbox{ $M=\sup_{x\in X} F(x)$} and $ m= \inf_{y \in X} F(y).$ By taking the supremum over elements $x$ and the infimum over $y$, the~Lipschitz condition implies that $M-m\leq 1.$ So
\begin{align}
  \int_{X} F d\mu - \int_{X} F d\nu  &=\sum_{x \in X} F(x)(f(x)-g(x)) \\
  &\leq \sum_{x: f(x) \geq g(x) } M(f(x)-g(x)) + \sum_{x: f(x) < g(x) } m(f(x)-g(x))\\
  &\leq \sum_{x: f(x) \geq g(x) } (m+1)(f(x)-g(x)) + \sum_{x: f(x) < g(x) } m(f(x)-g(x))\\
\label{eq:secondsummand}  &=\sum_{x: f(x) \geq g(x) } f(x)-g(x) + \sum_{x \in X}m(f(x)-g(x)) \\
  &=\sum_{x: f(x) \geq g(x) } f(x)-g(x),
\end{align}
\textls[-25]{using the fact that $\sum_x f(x)=\sum_x g(x)=1$ to eliminate the second summand of \mbox{Equation (\ref{eq:secondsummand})}. Next, we set}
\begin{align}
    P&=\sum_{x: f(x) \geq g(x) } f(x)-g(x),\\
    N&=\sum_{x: f(x) < g(x) } g(x)-f(x).
\end{align}

So $P-N = \sum_{x\in X} f(x)-g(x) = 0,$ whereas $P+N = \sum_{x \in X} |f(x)-g(x)|=\|f-g\|_1.$ It follows $P=N=\frac12 \|f-g\|_1,$ and $\left|\int_{X} F d\mu - \int_{X} F d\nu\right| \leq P$. Taking the supremum over $F$, we determine $W^1(f,g) \leq P$. Finally, let $F$ be the function defined in the proposition Statement (\ref{eq:bestF_new}). Then, $\int_{X} F d\mu - \int_{X} F d\nu$ coincides with $P$ by definition. Thus, the~supremal value $W^1(f,g)$ coincides exactly with $P$, and~$P=\frac12 \|f-g\|_1$, as~required.
\end{proof}

\begin{adjustwidth}{-\extralength}{0cm}
	
	\reftitle{References}

\begin{thebibliography}{999}
		
		\bibitem[Anandakugan(2022)]{coal_environment}
		Anandakugan, N.;
		\newblock {U.S. Energy Information Administration}. Coal Explained: Coal and the Environment. 2022.
		\newblock
		Available online: \url{https://www.eia.gov/energyexplained/coal/coal-and-the-environment.php} (accessed on 26 November 2022).
		
		\bibitem[Doenitz \em{et~al.}(1980)Doenitz, Schmidberger, Steinheil, and
		Streicher]{doenitz_hydrogen_1980}
		Doenitz, W.; Schmidberger, R.; Steinheil, E.; Streicher, R.
		\newblock Hydrogen production by high temperature electrolysis of water vapour.
		\newblock {\em Int. J. Hydrog. Energy} {\bf 1980}, {\em
			5},~55--63.
		\newblock {\url{https://doi.org/10.1016/0360-3199(80)90114-7}}.
		
		\bibitem[Wendt and Plzak(1991)]{wendt_hydrogen_1991}
		Wendt, H.; Plzak, V.
		\newblock Hydrogen production by water electrolysis / {Wasserstoffproduktion}
		durch {Elektrolyse} von {Wasser}.
		\newblock {\em Kerntechnik} {\bf 1991}, {\em 56},~22--28.
		\newblock {\url{https://doi.org/10.1515/kern-1991-560111}}.
		
		\bibitem[Khaselev(2001)]{khaselev_high-efficiency_2001}
		Khaselev, O.
		\newblock High-efficiency integrated multijunction photovoltaic/electrolysis
		systems for hydrogen production.
		\newblock {\em Int. J. Hydrog. Energy} {\bf 2001}, {\em
			26},~127--132.
		\newblock {\url{https://doi.org/10.1016/S0360-3199(00)00039-2}}.
		
		\bibitem[Ursua \em{et~al.}(2012)Ursua, Gandia, and
		Sanchis]{ursua_hydrogen_2012}
		Ursua, A.; Gandia, L.M.; Sanchis, P.
		\newblock Hydrogen {Production} {From} {Water} {Electrolysis}: {Current}
		{Status} and {Future} {Trends}.
		\newblock {\em Proc. IEEE} {\bf 2012}, {\em 100},~410--426.
		\newblock {\url{https://doi.org/10.1109/JPROC.2011.2156750}}.
		
		\bibitem[Chi and Yu(2018)]{chi_water_2018}
		Chi, J.; Yu, H.
		\newblock Water electrolysis based on renewable energy for hydrogen production.
		\newblock {\em Chin. J. Catal.} {\bf 2018}, {\em 39},~390--394.
		\newblock {\url{https://doi.org/10.1016/S1872-2067(17)62949-8}}.
		
		\bibitem[Zeng and Zhang(2010)]{zeng_recent_2010}
		Zeng, K.; Zhang, D.
		\newblock Recent progress in alkaline water electrolysis for hydrogen
		production and applications.
		\newblock {\em Prog. Energy Combust. Sci.} {\bf 2010}, {\em
			36},~307--326.
		\newblock {\url{https://doi.org/10.1016/j.pecs.2009.11.002}}.
		
		\bibitem[Barbir(2005)]{barbir_pem_2005}
		Barbir, F.
		\newblock {PEM} electrolysis for production of hydrogen from renewable energy
		sources.
		\newblock {\em Sol. Energy} {\bf 2005}, {\em 78},~661--669.
		\newblock {\url{https://doi.org/10.1016/j.solener.2004.09.003}}.
		
		\bibitem[Grigoriev \em{et~al.}(2006)Grigoriev, Porembsky, and
		Fateev]{grigoriev_pure_2006}
		Grigoriev, S.; Porembsky, V.; Fateev, V.
		\newblock Pure hydrogen production by {PEM} electrolysis for hydrogen energy.
		\newblock {\em Int. J. Hydrog. Energy} {\bf 2006}, {\em
			31},~171--175.
		\newblock {\url{https://doi.org/10.1016/j.ijhydene.2005.04.038}}.
		
		\bibitem[Bessarabov \em{et~al.}(2016)Bessarabov, Wang, Li, and
		Zhao]{bessarabov_pem_2016}
		Bessarabov, D.G.; Wang, H.; Li, H.; Zhao, N., (Eds.)
		\newblock {\em {PEM} Electrolysis for Hydrogen Production: Principles and
			Applications}; CRC Press: Boca Raton, FL, USA, 2016.
		
		\bibitem[Guo \em{et~al.}(2019)Guo, Li, Zhou, and Liu]{guo_comparison_2019}
		Guo, Y.; Li, G.; Zhou, J.; Liu, Y.
		\newblock Comparison between hydrogen production by alkaline water electrolysis
		and hydrogen production by {PEM} electrolysis.
		\newblock {\em IOP Conf. Ser. Earth Environ. Sci.} {\bf
			2019}, {\em 371},~042022.
		\newblock {\url{https://doi.org/10.1088/1755-1315/371/4/042022}}.
		
		\bibitem[Yilmaz and Kanoglu(2014)]{yilmaz_thermodynamic_2014}
		Yilmaz, C.; Kanoglu, M.
		\newblock Thermodynamic evaluation of geothermal energy powered hydrogen
		production by {PEM} water electrolysis.
		\newblock {\em Energy} {\bf 2014}, {\em 69},~592--602.
		\newblock {\url{https://doi.org/10.1016/j.energy.2014.03.054}}.
		
		\bibitem[Rozendal \em{et~al.}(2006)Rozendal, Hamelers, Euverink, Metz, and
		Buisman]{rozendal_principle_2006}
		Rozendal, R.; Hamelers, H.; Euverink, G.; Metz, S.; Buisman, C.
		\newblock Principle and perspectives of hydrogen production through
		biocatalyzed electrolysis.
		\newblock {\em Int. J. Hydrog. Energy} {\bf 2006}, {\em
			31},~1632--1640.
		\newblock {\url{https://doi.org/10.1016/j.ijhydene.2005.12.006}}.
		
		\bibitem[Call and Logan(2008)]{call_hydrogen_2008}
		Call, D.; Logan, B.E.
		\newblock Hydrogen {Production} in a {Single} {Chamber} {Microbial}
		{Electrolysis} {Cell} {Lacking} a {Membrane}.
		\newblock {\em Environ. Sci. Technol.} {\bf 2008}, {\em
			42},~3401--3406.
		\newblock {\url{https://doi.org/10.1021/es8001822}}.
		
		\bibitem[Rai \em{et~al.}(2022)Rai, Bhui, and V]{Rai2022}
		\textls[-15]{Rai, C.; Bhui, B.; V, P.
			\newblock Techno-economic analysis of e-waste based chemical looping reformer
			as hydrogen generator with co-generation of metals, electricity and syngas.
			\newblock {\em Int. J. Hydrog. Energy} {\bf 2022}, {\em
				47},~11177--11189.
			\newblock {\url{https://doi.org/10.1016/j.ijhydene.2022.01.159}}.}
		
		\bibitem[Wang \em{et~al.}(2014)Wang, Wang, Gong, and
		Guo]{wang_intensification_2014}
		Wang, M.; Wang, Z.; Gong, X.; Guo, Z.
		\newblock The intensification technologies to water electrolysis for hydrogen
		production---A review.
		\newblock {\em Renew. Sustain. Energy Rev.} {\bf 2014}, {\em
			29},~573--588.
		\newblock {\url{https://doi.org/10.1016/j.rser.2013.08.090}}.
		
		\bibitem[Chakik \em{et~al.}(2017)Chakik, Kaddami, and
		Mikou]{chakik_effect_2017}
		Chakik, F.E.; Kaddami, M.; Mikou, M.
		\newblock Effect of operating parameters on hydrogen production by electrolysis
		of water.
		\newblock {\em Int. J. Hydrog. Energy} {\bf 2017}, {\em
			42},~25550--25557.
		\newblock {\url{https://doi.org/10.1016/j.ijhydene.2017.07.015}}.
		
		\bibitem[Wu \em{et~al.}(2022)Wu, Zhou, Xiao, Xu, and Li]{Wu2022}
		Wu, L.; Zhou, Z.; Xiao, Y.; Xu, Z.; Li, X.
		\newblock Hydrogen evolution reaction activity and stability of sintered porous
		{Ni-Cu-Ti-La}$_2$O$_3$ cathodes in a wide {pH} range.
		\newblock {\em Int. J. Hydrog. Energy} {\bf 2022}, {\em
			47},~11101--11115.
		\newblock {\url{https://doi.org/10.1016/j.ijhydene.2022.01.019}}.
		
		\bibitem[Ikeda \em{et~al.}(2022)Ikeda, Misumi, Kojima, Haleem, Kuroda, and
		Mitsushima]{Ikeda2022}
		Ikeda, H.; Misumi, R.; Kojima, Y.; Haleem, A.A.; Kuroda, Y.; Mitsushima, S.
		\newblock Microscopic high-speed video observation of oxygen bubble generation
		behavior and effects of anode electrode shape on {OER} performance in
		alkaline water electrolysis.
		\newblock {\em Int. J. Hydrog. Energy} {\bf 2022}, {\em
			47},~11116--11127.
		\newblock {\url{https://doi.org/10.1016/j.ijhydene.2022.01.166}}.
		
		\bibitem[Zhang \em{et~al.}(2010)Zhang, Lin, and Chen]{zhang_evaluation_2010}
		Zhang, H.; Lin, G.; Chen, J.
		\newblock Evaluation and calculation on the efficiency of a water electrolysis
		system for hydrogen production.
		\newblock {\em Int. J. Hydrog. Energy} {\bf 2010}, {\em
			35},~10851--10858.
		\newblock {\url{https://doi.org/10.1016/j.ijhydene.2010.07.088}}.
		
		\bibitem[Shit \em{et~al.}(2022)Shit, Bolar, Kolya, Kang, Murmu, and
		Kuila]{Shit2022}
		Shit, S.; Bolar, S.; Kolya, H.; Kang, C.W.; Murmu, N.C.; Kuila, T.
		\newblock Assisting the formation of {S}-doped {FeMoO}$_4$ in lieu of an iron
		oxide/molybdenum sulfide heterostructure: A unique approach towards attaining
		excellent electrocatalytic water splitting activity.
		\newblock {\em Int. J. Hydrog. Energy} {\bf 2022}, {\em
			47},~11128--11142.
		\newblock {\url{https://doi.org/10.1016/j.ijhydene.2022.01.165}}.
		
		\bibitem[Tong \em{et~al.}(2022)Tong, Liu, Song, Zhang, Latthe, Liu, and
		Xing]{Tong2022}
		\textls[-15]{Tong, L.; Liu, Y.; Song, C.; Zhang, Y.; Latthe, S.S.; Liu, S.; Xing, R.
			\newblock {(Fe, Ni)S}$_2$@{MoS}$_2$/{NiS}$_2$ hollow heterostructure nanocubes
			for high-performance alkaline water electrolysis.
			\newblock {\em Int. J. Hydrog. Energy} {\bf 2022}, {\em
				47},~11143--11152.
			\newblock {\url{https://doi.org/10.1016/j.ijhydene.2022.01.161}}.}
		
		\bibitem[Liang \em{et~al.}(2022)Liang, Sui, Li, Guo, Luo, Xu, Yao, Wang, and
		Chen]{Liang2022}
		Liang, S.; Sui, G.; Li, J.; Guo, D.; Luo, Z.; Xu, R.; Yao, H.; Wang, C.; Chen,
		S.
		\newblock {ZIF}-L-derived porous C-doped {ZnO}/{CdS} graded nanorods with
		Z-scheme heterojunctions for enhanced photocatalytic hydrogen evolution.
		\newblock {\em Int. J. Hydrog. Energy} {\bf 2022}, {\em
			47},~11190--11202.
		\newblock {\url{https://doi.org/10.1016/j.ijhydene.2022.01.154}}.
		
		\bibitem[Khasanah \em{et~al.}(2022)Khasanah, Lin, Ho, Peng, Hsiao, Wang, and
		Chien]{Khasanah2022}
		Khasanah, R.A.N.; Lin, H.C.; Ho, H.Y.; Peng, Y.P.; Hsiao, H.L.; Wang, C.R.;
		Chien, F.S.S.
		\newblock Photoelectrocatalytic hydrolysis of ammonia borane by electrochemical
		deposited cuprous oxide on titanium dioxide nanotube arrays.
		\newblock {\em Int. J. Hydrog. Energy} {\bf 2022}, {\em
			47},~11203--11210.
		\newblock {\url{https://doi.org/10.1016/j.ijhydene.2022.01.167}}.
		
		\bibitem[Liu \em{et~al.}(2022)Liu, Shi, Liu, and Li]{Liu2022_hydrogen}
		Liu, Y.; Shi, X.; Liu, X.; Li, X.
		\newblock Facile synthesis of {C-Ta}$^{4+}$ co-doped {NaTaO}$_3$ and {rGO}
		nanocomposites with enhanced visible light photocatalytic performance.
		\newblock {\em Int. J. Hydrog. Energy} {\bf 2022}, {\em
			47},~11211--11223.
		\newblock {\url{https://doi.org/10.1016/j.ijhydene.2022.01.163}}.
		
		\bibitem[Ulate-Kolitsky \em{et~al.}(2022)Ulate-Kolitsky, Tougas, and
		Huot]{UlateKolitsky2022}
		Ulate-Kolitsky, E.; Tougas, B.; Huot, J.
		\newblock First Hydrogenation of {TiFe} with Addition of 20 wt.{\%} {T}i.
		\newblock {\em Hydrogen} {\bf 2022}, {\em 3},~379--388.
		\newblock {\url{https://doi.org/10.3390/hydrogen3040023}}.
		
		\bibitem[He \em{et~al.}(2022)He, Han, Xiang, Ran, Wang, Zhou, and
		Shao]{He2022_hydrogen}
		He, J.; Han, X.; Xiang, H.; Ran, R.; Wang, W.; Zhou, W.; Shao, Z.
		\newblock Aluminum Cation Doping in {R}uddlesden-{P}opper {Sr2TiO4} Enables
		High-Performance Photocatalytic Hydrogen Evolution.
		\newblock {\em Hydrogen} {\bf 2022}, {\em 3},~501--511.
		\newblock {\url{https://doi.org/10.3390/hydrogen3040032}}.
		
		\bibitem[So{\l}owski \em{et~al.}(2022)So{\l}owski, Shalaby, and
		\"{O}zdemir]{Soowski2022}
		So{\l}owski, G.; Shalaby, M.; \"{O}zdemir, F.A.
		\newblock Plastic and Waste Tire Pyrolysis Focused on Hydrogen Production---A
		Review.
		\newblock {\em Hydrogen} {\bf 2022}, {\em 3},~531--549.
		\newblock {\url{https://doi.org/10.3390/hydrogen3040034}}.
		
		\bibitem[Shelepova \em{et~al.}(2022)Shelepova, Maksimova, Bauman, Mishakov, and
		Vedyagin]{Shelepova2022}
		Shelepova, E.V.; Maksimova, T.A.; Bauman, Y.I.; Mishakov, I.V.; Vedyagin, A.A.
		\newblock Experimental and Simulation Study on Coproduction of Hydrogen and
		Carbon Nanomaterials by Catalytic Decomposition of Methane-Hydrogen Mixtures.
		\newblock {\em Hydrogen} {\bf 2022}, {\em 3},~450--462.
		\newblock {\url{https://doi.org/10.3390/hydrogen3040028}}.
		
		\bibitem[Vedyagin \em{et~al.}(2021)Vedyagin, Mishakov, Korneev, Bauman,
		Nalivaiko, and Gromov]{Vedyagin2021}
		Vedyagin, A.A.; Mishakov, I.V.; Korneev, D.V.; Bauman, Y.I.; Nalivaiko, A.Y.;
		Gromov, A.A.
		\newblock Selected Aspects of Hydrogen Production via Catalytic Decomposition
		of Hydrocarbons.
		\newblock {\em Hydrogen} {\bf 2021}, {\em 2},~122--133.
		\newblock {\url{https://doi.org/10.3390/hydrogen2010007}}.
		
		\bibitem[Lys \em{et~al.}(2020)Lys, Fadonougbo, Faisal, Suh, Lee, Shim, Park,
		and Cho]{Lys2020_hydrogen}
		Lys, A.; Fadonougbo, J.O.; Faisal, M.; Suh, J.Y.; Lee, Y.S.; Shim, J.H.; Park,
		J.; Cho, Y.W.
		\newblock Enhancing the Hydrogen Storage Properties of {AxBy} Intermetallic
		Compounds by Partial Substitution: A Short Review.
		\newblock {\em Hydrogen} {\bf 2020}, {\em 1},~38--63.
		\newblock {\url{https://doi.org/10.3390/hydrogen1010004}}.
		
		\bibitem[Heinemann \em{et~al.}(2022)Heinemann, Wilkinson, Adie, Edlmann,
		Thaysen, Hassanpouryouzband, and Haszeldine]{Heinemann2022}
		Heinemann, N.; Wilkinson, M.; Adie, K.; Edlmann, K.; Thaysen, E.M.;
		Hassanpouryouzband, A.; Haszeldine, R.S.
		\newblock Cushion Gas in Hydrogen Storage---A Costly {CAPEX} or a Valuable
		Resource for Energy Crises?
		\newblock {\em Hydrogen} {\bf 2022}, {\em 3},~550--563.
		\newblock {\url{https://doi.org/10.3390/hydrogen3040035}}.
		
		\bibitem[Pistidda(2021)]{Pistidda2021}
		Pistidda, C.
		\newblock Solid-State Hydrogen Storage for a Decarbonized Society.
		\newblock {\em Hydrogen} {\bf 2021}, {\em 2},~428--443.
		\newblock {\url{https://doi.org/10.3390/hydrogen2040024}}.
		
		\bibitem[Ekhtiari \em{et~al.}(2022)Ekhtiari, Flynn, and Syron]{Ekhtiari2022}
		Ekhtiari, A.; Flynn, D.; Syron, E.
		\newblock Green Hydrogen Blends with Natural Gas and Its Impact on the Gas
		Network.
		\newblock {\em Hydrogen} {\bf 2022}, {\em 3},~402--417.
		\newblock {\url{https://doi.org/10.3390/hydrogen3040025}}.
		
		\bibitem[Lattin and Utgikar(2007)]{lattin_transition_2007}
		Lattin, W.; Utgikar, V.
		\newblock Transition to hydrogen economy in the {United} {States}: {A} 2006
		status report.
		\newblock {\em Int. J. Hydrog. Energy} {\bf 2007}, {\em
			32},~3230--3237.
		\newblock {\url{https://doi.org/10.1016/j.ijhydene.2007.02.004}}.
		
		\bibitem[Park(2013)]{park_country-dependent_2013}
		Park, S.
		\newblock The country-dependent shaping of `hydrogen niche' formation: {A}
		comparative case study of the {UK} and {South} {Korea} from the innovation
		system perspective.
		\newblock {\em Int. J. Hydrog. Energy} {\bf 2013}, {\em
			38},~6557--6568.
		\newblock {\url{https://doi.org/10.1016/j.ijhydene.2013.03.114}}.
		
		\bibitem[Yuan and Lin(2010)]{yuan_hydrogen_2010}
		Yuan, K.; Lin, W.
		\newblock Hydrogen in {China}: {Policy}, program and progress.
		\newblock {\em Int. J. Hydrog. Energy} {\bf 2010}, {\em
			35},~3110--3113.
		\newblock {\url{https://doi.org/10.1016/j.ijhydene.2009.08.042}}.
		
		\bibitem[Collera and Agaton(2021)]{collera_opportunities_2021}
		Collera, A.A.; Agaton, C.B.
		\newblock Opportunities for production and utilization of green hydrogen in the
		{P}hilippines.
		\newblock {\em Int. J. Energy Econ. Policy} {\bf
			2021}, {\em 11},~37--41.
		\newblock {\url{https://doi.org/10.32479/ijeep.11383}}.
		
		\bibitem[Ramirez-Salgado and
		Estrada-Martinez(2004)]{ramirez-salgado_roadmap_2004}
		Ramirez-Salgado, J.; Estrada-Martinez, A.
		\newblock Roadmap towards a sustainable hydrogen economy in {Mexico}.
		\newblock {\em J. Power Sources} {\bf 2004}, {\em 129},~255--263.
		\newblock {\url{https://doi.org/10.1016/j.jpowsour.2003.11.054}}.
		
		\bibitem[Touili \em{et~al.}(2018)Touili, Alami~Merrouni, Azouzoute,
		El~Hassouani, and Amrani]{touili_technical_2018}
		Touili, S.; Alami~Merrouni, A.; Azouzoute, A.; El~Hassouani, Y.; Amrani, A.i.
		\newblock A technical and economical assessment of hydrogen production
		potential from solar energy in {Morocco}.
		\newblock {\em Int. J. Hydrog. Energy} {\bf 2018}, {\em
			43},~22777--22796.
		\newblock {\url{https://doi.org/10.1016/j.ijhydene.2018.10.136}}.
		
		\bibitem[Apak \em{et~al.}(2012)Apak, Atay, and Tuncer]{apak_renewable_2012}
		Apak, S.; Atay, E.; Tuncer, G.
		\newblock Renewable hydrogen energy regulations, codes and standards:
		{Challenges} faced by an {EU} candidate country.
		\newblock {\em Int. J. Hydrog. Energy} {\bf 2012}, {\em
			37},~5481--5497.
		\newblock {\url{https://doi.org/10.1016/j.ijhydene.2012.01.005}}.
		
		\bibitem[James and Menzies(2022)]{james2021_hydrogen}
		James, N.; Menzies, M.
		\newblock Spatio-temporal trends in the propagation and capacity of low-carbon
		hydrogen projects.
		\newblock {\em Int. J. Hydrog. Energy} {\bf 2022}, {\em
			47},~16775--16784.
		\newblock {\url{https://doi.org/10.1016/j.ijhydene.2022.03.198}}.
		
		\bibitem[Bridgeland \em{et~al.}(2022)Bridgeland, Chapman, McLellan, Sofronis,
		and Fujii]{Bridgeland2022}
		\textls[-30]{Bridgeland, R.; Chapman, A.; McLellan, B.; Sofronis, P.; Fujii, Y.
			\newblock Challenges toward achieving a successful hydrogen economy in the
			{US}: Potential end-use and infrastructure analysis to the year 2100.
			\newblock {\em Clean. Prod. Lett.} {\bf 2022}, {\em 3},~100012.
			\newblock {\url{https://doi.org/10.1016/j.clpl.2022.100012}}.}
		
		\bibitem[Saeedmanesh \em{et~al.}(2018)Saeedmanesh, Kinnon, and
		Brouwer]{Saeedmanesh2018}
		Saeedmanesh, A.; Kinnon, M.A.M.; Brouwer, J.
		\newblock Hydrogen is essential for sustainability.
		\newblock {\em Curr. Opin. Electrochem.} {\bf 2018}, {\em
			12},~166--181.
		\newblock {\url{https://doi.org/10.1016/j.coelec.2018.11.009}}.
		
		\bibitem[Dawood \em{et~al.}(2020)Dawood, Anda, and Shafiullah]{Dawood2020}
		Dawood, F.; Anda, M.; Shafiullah, G.
		\newblock Hydrogen production for energy: An overview.
		\newblock {\em Int. J. Hydrog. Energy} {\bf 2020}, {\em
			45},~3847--3869.
		\newblock {\url{https://doi.org/10.1016/j.ijhydene.2019.12.059}}.
		
		\bibitem[Pleshivtseva \em{et~al.}(2023)Pleshivtseva, Derevyanov, Pimenov, and
		Rapoport]{Pleshivtseva2023}
		Pleshivtseva, Y.; Derevyanov, M.; Pimenov, A.; Rapoport, A.
		\newblock Comprehensive review of low carbon hydrogen projects towards the
		decarbonization pathway.
		\newblock {\em Int. J. Hydrog. Energy} {\bf 2023}, {\em
			48},~3703--3724.
		\newblock {\url{https://doi.org/10.1016/j.ijhydene.2022.10.209}}.
		
		\bibitem[Manchein \em{et~al.}(2020)Manchein, Brugnago, da~Silva, Mendes, and
		Beims]{Manchein2020}
		Manchein, C.; Brugnago, E.L.; da~Silva, R.M.; Mendes, C.F.O.; Beims, M.W.
		\newblock Strong correlations between power-law growth of {COVID}-19 in four
		continents and the inefficiency of soft quarantine strategies.
		\newblock {\em Chaos Interdiscip. J. Nonlinear Sci.} {\bf
			2020}, {\em 30},~041102.
		\newblock {\url{https://doi.org/10.1063/5.0009454}}.
		
		\bibitem[James \em{et~al.}(2022)James, Menzies, and
		Bondell]{james2021_CovidIndia}
		James, N.; Menzies, M.; Bondell, H.
		\newblock Comparing the dynamics of {COVID}-19 infection and mortality in the
		{U}nited {S}tates, {I}ndia, and {B}razil.
		\newblock {\em Phys. D Nonlinear Phenom.} {\bf 2022}, {\em 432},~133158.
		\newblock {\url{https://doi.org/10.1016/j.physd.2022.133158}}.
		
		\bibitem[Li \em{et~al.}(2021)Li, Xu, Song, Wang, and Perc]{Li2021_Matjaz}
		Li, H.J.; Xu, W.; Song, S.; Wang, W.X.; Perc, M.
		\newblock The dynamics of epidemic spreading on signed networks.
		\newblock {\em Chaos Solitons Fractals} {\bf 2021}, {\em 151},~111294.
		\newblock {\url{https://doi.org/10.1016/j.chaos.2021.111294}}.
		
		\bibitem[Blasius(2020)]{Blasius2020}
		Blasius, B.
		\newblock Power-law distribution in the number of confirmed {COVID}-19 cases.
		\newblock {\em Chaos Interdiscip. J. Nonlinear Sci.} {\bf
			2020}, {\em 30},~093123.
		\newblock {\url{https://doi.org/10.1063/5.0013031}}.
		
		\bibitem[James and Menzies(2022)]{james2021_TVO}
		James, N.; Menzies, M.
		\newblock Estimating a continuously varying offset between multivariate time
		series with application to {COVID}-19 in the {U}nited {S}tates.
		\newblock {\em  Eur. Phys. J. Spec. Top.} {\bf 2022}, {\em
			231},~3419--3426.
		\newblock {\url{https://doi.org/10.1140/epjs/s11734-022-00430-y}}.
		
		\bibitem[Perc \em{et~al.}(2020)Perc, Miksi{\'{c}}, Slavinec, and
		Sto{\v{z}}er]{Perc2020}
		\textls[-25]{Perc, M.; Miksi{\'{c}}, N.G.; Slavinec, M.; Sto{\v{z}}er, A.
			\newblock Forecasting {COVID}-19.
			\newblock {\em Front. Phys.} {\bf 2020}, {\em 8},~127.
			\newblock {\url{https://doi.org/10.3389/fphy.2020.00127}}}.
		
		\bibitem[Machado and Lopes(2020)]{Machado2020}
		Machado, J.A.T.; Lopes, A.M.
		\newblock Rare and extreme events: The case of {COVID}-19 pandemic.
		\newblock {\em Nonlinear Dyn.} {\bf 2020}, \emph{100}, 2953--2972. 
		\newblock {\url{https://doi.org/10.1007/s11071-020-05680-w}}.
		
		\bibitem[James and Menzies(2022)]{james2022_CO2}
		James, N.; Menzies, M.
		\newblock Global and regional changes in carbon dioxide emissions: 1970--2019.
		\newblock {\em Phys. A Stat. Mech. Appl.} {\bf
			2022}, {\em 608},~128302.
		\newblock {\url{https://doi.org/10.1016/j.physa.2022.128302}}.
		
		\bibitem[Khan \em{et~al.}(2020)Khan, Khan, and Rehan]{Khan2020}
		Khan, M.K.; Khan, M.I.; Rehan, M.
		\newblock The relationship between energy consumption, economic growth and
		carbon dioxide emissions in Pakistan.
		\newblock {\em Financ. Innov.} {\bf 2020}, {\em 6}, 1.
		\newblock {\url{https://doi.org/10.1186/s40854-019-0162-0}}.

  		\bibitem[James and Menzies(2023)]{james2020_LP}
		James, N.; Menzies, M.
		\newblock Equivalence relations and $L^p$ distances between time series with application to the Black Summer Australian bushfires.
		\newblock {\em Phys. D Nonlinear Phenom.} {\bf 2023}, {\em 448},~133693.
		\newblock {\url{https://doi.org/10.1016/j.physd.2023.133693}}.
		
		\bibitem[Dro{\.{z}}d{\.{z}} \em{et~al.}(2021)Dro{\.{z}}d{\.{z}}, Kwapie{\'{n}},
		and O{\'{s}}wi{\k{e}}cimka]{Drod2021_entropy}
		Dro{\.{z}}d{\.{z}}, S.; Kwapie{\'{n}}, J.; O{\'{s}}wi{\k{e}}cimka, P.
		\newblock Complexity in Economic and Social Systems.
		\newblock {\em Entropy} {\bf 2021}, {\em 23},~133.
		\newblock {\url{https://doi.org/10.3390/e23020133}}.
		
		\bibitem[James \em{et~al.}(2022)James, Menzies, and Gottwald]{james_georg}
		James, N.; Menzies, M.; Gottwald, G.A.
		\newblock On financial market correlation structures and diversification
		benefits across and within equity sectors.
		\newblock {\em Phys. A Stat. Mech. Appl.} {\bf
			2022}, {\em 604},~127682.
		\newblock {\url{https://doi.org/10.1016/j.physa.2022.127682}}.
		
		\bibitem[Liu \em{et~al.}(1997)Liu, Cizeau, Meyer, Peng, and Stanley]{Liu1997}
		Liu, Y.; Cizeau, P.; Meyer, M.; Peng, C.K.; Stanley, H.E.
		\newblock Correlations in economic time series.
		\newblock {\em Phys. A Stat. Mech. Appl.} {\bf
			1997}, {\em 245},~437--440.
		\newblock {\url{https://doi.org/10.1016/s0378-4371(97)00368-3}}.
		
		\bibitem[Basalto \em{et~al.}(2007)Basalto, Bellotti, Carlo, Facchi, Pantaleo,
		and Pascazio]{Basalto2007}
		Basalto, N.; Bellotti, R.; Carlo, F.D.; Facchi, P.; Pantaleo, E.; Pascazio, S.
		\newblock Hausdorff clustering of financial time series.
		\newblock {\em Phys. A Stat. Mech. Appl.} {\bf
			2007}, {\em 379},~635--644.
		\newblock {\url{https://doi.org/10.1016/j.physa.2007.01.011}}.
		
		\bibitem[W{\k{a}}torek \em{et~al.}(2021)W{\k{a}}torek, Kwapie{\'{n}}, and
		Dro{\.{z}}d{\.{z}}]{Wtorek2021_entropy}
		W{\k{a}}torek, M.; Kwapie{\'{n}}, J.; Dro{\.{z}}d{\.{z}}, S.
		\newblock Financial Return Distributions: Past, Present, and {COVID}-19.
		\newblock {\em Entropy} {\bf 2021}, {\em 23},~884.
		\newblock {\url{https://doi.org/10.3390/e23070884}}.
		
		\bibitem[Prakash \em{et~al.}(2021)Prakash, James, Menzies, and
		Francis]{james_arjun}
		Prakash, A.; James, N.; Menzies, M.; Francis, G.
		\newblock Structural Clustering of Volatility Regimes for Dynamic Trading
		Strategies.
		\newblock {\em Appl. Math. Financ.} {\bf 2021}, {\em 28},~236--274.
		\newblock {\url{https://doi.org/10.1080/1350486x.2021.2007146}}.
		
		\bibitem[Dro{\.{z}}d{\.{z}} \em{et~al.}(2001)Dro{\.{z}}d{\.{z}}, Gr\"{u}mmer,
		Ruf, and Speth]{Drod2001}
		Dro{\.{z}}d{\.{z}}, S.; Gr\"{u}mmer, F.; Ruf, F.; Speth, J.
		\newblock Towards identifying the world stock market cross-correlations: {DAX}
		versus {D}ow {J}ones.
		\newblock {\em Phys. A Stat. Mech. Appl.} {\bf
			2001}, {\em 294},~226--234.
		\newblock {\url{https://doi.org/10.1016/s0378-4371(01)00119-4}}.
		
		\bibitem[James \em{et~al.}(2022)James, Menzies, and
		Chin]{james2022_stagflation}
		James, N.; Menzies, M.; Chin, K.
		\newblock Economic state classification and portfolio optimisation with
		application to stagflationary environments.
		\newblock {\em Chaos Solitons Fractals} {\bf 2022}, {\em 164},~112664.
		\newblock {\url{https://doi.org/10.1016/j.chaos.2022.112664}}.
		
		\bibitem[G{\k{e}}barowski \em{et~al.}(2019)G{\k{e}}barowski,
		O{\'{s}}wi{\k{e}}cimka, W{\k{a}}torek, and Dro{\.{z}}d{\.{z}}]{Gbarowski2019}
		\textls[-35]{G{\k{e}}barowski, R.; O{\'{s}}wi{\k{e}}cimka, P.; W{\k{a}}torek, M.;
			Dro{\.{z}}d{\.{z}}, S.
			\newblock Detecting correlations and triangular arbitrage opportunities in the
			Forex by means of multifractal detrended cross-correlations analysis.
			\newblock {\em Nonlinear Dyn.} {\bf 2019}, {\em 98},~2349--2364.
			\newblock {\url{https://doi.org/10.1007/s11071-019-05335-5}}.}
		
		\bibitem[James and Menzies(2021)]{james2021_MJW}
		James, N.; Menzies, M.
		\newblock A new measure between sets of probability distributions with
		applications to erratic financial behavior.
		\newblock {\em J. Stat. Mech. Theory Exp.} {\bf
			2021}, {\em 2021},~123404.
		\newblock {\url{https://doi.org/10.1088/1742-5468/ac3d91}}.
		
		\bibitem[Sigaki \em{et~al.}(2019)Sigaki, Perc, and Ribeiro]{Sigaki2019}
		Sigaki, H.Y.D.; Perc, M.; Ribeiro, H.V.
		\newblock Clustering patterns in efficiency and the coming-of-age of the
		cryptocurrency market.
		\newblock {\em Sci. Rep.} {\bf 2019}, {\em 9},~1440.
		\newblock {\url{https://doi.org/10.1038/s41598-018-37773-3}}.
		
		\bibitem[Dro{\.{z}}d{\.{z}} \em{et~al.}(2020)Dro{\.{z}}d{\.{z}}, Kwapie{\'{n}},
		O{\'{s}}wi{\k{e}}cimka, Stanisz, and W{\k{a}}torek]{Drod2020_entropy}
		Dro{\.{z}}d{\.{z}}, S.; Kwapie{\'{n}}, J.; O{\'{s}}wi{\k{e}}cimka, P.; Stanisz,
		T.; W{\k{a}}torek, M.
		\newblock Complexity in Economic and Social Systems: Cryptocurrency Market at
		around {COVID}-19.
		\newblock {\em Entropy} {\bf 2020}, {\em 22},~1043.
		\newblock {\url{https://doi.org/10.3390/e22091043}}.
		
		\bibitem[James and Menzies(2022)]{James2021_crypto2}
		James, N.; Menzies, M.
		\newblock Collective correlations, dynamics, and behavioural inconsistencies of
		the cryptocurrency market over time.
		\newblock {\em Nonlinear Dyn.} {\bf 2022}, {\em 107},~4001--4017.
		\newblock {\url{https://doi.org/10.1007/s11071-021-07166-9}}.
		
		\bibitem[Dro{\.{z}}d{\.{z}} \em{et~al.}(2020)Dro{\.{z}}d{\.{z}}, Minati,
		O{\'{s}}wi{\k{e}}cimka, Stanuszek, and W{\k{a}}torek]{Drod2020}
		Dro{\.{z}}d{\.{z}}, S.; Minati, L.; O{\'{s}}wi{\k{e}}cimka, P.; Stanuszek, M.;
		W{\k{a}}torek, M.
		\newblock Competition of noise and collectivity in global cryptocurrency
		trading: Route to a self-contained market.
		\newblock {\em Chaos: Interdiscip. J. Nonlinear Sci.} {\bf
			2020}, {\em 30},~023122.
		\newblock {\url{https://doi.org/10.1063/1.5139634}}.
		
		\bibitem[W{\k{a}}torek \em{et~al.}(2021)W{\k{a}}torek, Dro{\.{z}}d{\.{z}},
		Kwapie{\'{n}}, Minati, O{\'{s}}wi{\k{e}}cimka, and Stanuszek]{Wtorek2020}
		W{\k{a}}torek, M.; Dro{\.{z}}d{\.{z}}, S.; Kwapie{\'{n}}, J.; Minati, L.;
		O{\'{s}}wi{\k{e}}cimka, P.; Stanuszek, M.
		\newblock Multiscale characteristics of the emerging global cryptocurrency
		market.
		\newblock {\em Phys. Rep.} {\bf 2021}, {\em 901},~1--82.
		\newblock {\url{https://doi.org/10.1016/j.physrep.2020.10.005}}.
		
		\bibitem[James and Menzies(2022)]{james2022_guns}
		James, N.; Menzies, M.
		\newblock Dual-domain analysis of gun violence incidents in the {U}nited
		{S}tates.
		\newblock {\em Chaos Interdiscip. J. Nonlinear Sci.} {\bf
			2022}, {\em 32},~111101.
		\newblock {\url{https://doi.org/10.1063/5.0120822}}.
		
		\bibitem[Perc \em{et~al.}(2013)Perc, Donnay, and Helbing]{Perc2013}
		Perc, M.; Donnay, K.; Helbing, D.
		\newblock Understanding Recurrent Crime as System-Immanent Collective Behavior.
		\newblock {\em {PLoS} {ONE}} {\bf 2013}, {\em 8},~e76063.
		\newblock {\url{https://doi.org/10.1371/journal.pone.0076063}}.
		
		\bibitem[James \em{et~al.}(2023)James, Menzies, Chok, Milner, and
		Milner]{James2023_terrorist}
		James, N.; Menzies, M.; Chok, J.; Milner, A.; Milner, C.
		\newblock Geometric persistence and distributional trends in worldwide
		terrorism.
		\newblock {\em Chaos Solitons Fractals} {\bf 2023}, {\em 169},~113277.
		\newblock {\url{https://doi.org/10.1016/j.chaos.2023.113277}}.
		
		\bibitem[Ribeiro \em{et~al.}(2012)Ribeiro, Mukherjee, and Zeng]{Ribeiro2012}
		Ribeiro, H.V.; Mukherjee, S.; Zeng, X.H.T.
		\newblock Anomalous diffusion and long-range correlations in the score
		evolution of the game of cricket.
		\newblock {\em Phys. Rev. E} {\bf 2012}, {\em 86}, 022102.
		\newblock {\url{https://doi.org/10.1103/physreve.86.022102}}.
		
		\bibitem[James and Menzies(2022)]{james2021_spectral}
		James, N.; Menzies, M.
		\newblock Optimally adaptive {B}ayesian spectral density estimation for
		stationary and nonstationary processes.
		\newblock {\em Stat. Comput.} {\bf 2022}, {\em 32},~45.
		\newblock {\url{https://doi.org/10.1007/s11222-022-10103-4}}.
		
		\bibitem[Merritt and Clauset(2014)]{Merritt2014}
		Merritt, S.; Clauset, A.
		\newblock Scoring dynamics across professional team sports: Tempo, balance and
		predictability.
		\newblock {\em {EPJ} Data Sci.} {\bf 2014}, {\em 3
		}, 4.
		\newblock {\url{https://doi.org/10.1140/epjds29}}.
		
		\bibitem[James \em{et~al.}(2022)James, Menzies, and
		Bondell]{james2021_olympics}
		James, N.; Menzies, M.; Bondell, H.
		\newblock In search of peak human athletic potential: A mathematical
		investigation.
		\newblock {\em Chaos Interdiscip. J. Nonlinear Sci.} {\bf
			2022}, {\em 32},~023110.
		\newblock {\url{https://doi.org/10.1063/5.0073141}}.
		
		\bibitem[Clauset \em{et~al.}(2015)Clauset, Kogan, and Redner]{Clauset2015}
		Clauset, A.; Kogan, M.; Redner, S.
		\newblock Safe leads and lead changes in competitive team sports.
		\newblock {\em Phys. Rev. E} {\bf 2015}, {\em 91}, 062815. 
		\newblock {\url{https://doi.org/10.1103/physreve.91.062815}}.
		
		\bibitem[Hyd(2022)]{Hydrogendata}
		{International Energy Agency}. Hydrogen Projects Database. 2022.
		\newblock
		Available online: \url{https://www.iea.org/data-and-statistics/data-product/hydrogen-projects-database}  (accessed on 1 November 2022). 
		
		\bibitem[Ward(1963)]{Ward1963}
		Ward, J.H.
		\newblock Hierarchical Grouping to Optimize an Objective Function.
		\newblock {\em J. {A}m. Stat. Assoc.} {\bf 1963},
		{\em 58},~236--244.
		\newblock {\url{https://doi.org/10.1080/01621459.1963.10500845}}.
		
		\bibitem[Szekely and Rizzo(2005)]{Szekely2005}
		Szekely, G.J.; Rizzo, M.L.
		\newblock Hierarchical Clustering via Joint Between-Within Distances: Extending
		{W}ard's Minimum Variance Method.
		\newblock {\em J. Classif.} {\bf 2005}, {\em 22},~151--183.
		\newblock {\url{https://doi.org/10.1007/s00357-005-0012-9}}.
		
		\bibitem[M\"{u}llner(2013)]{Mllner2013}
		M\"{u}llner, D.
		\newblock Fastcluster: Fast Hierarchical, Agglomerative Clustering Routines
		{forRandPython}.
		\newblock {\em J. Stat. Softw.} {\bf 2013}, {\em 53}, 1--18.
		\newblock {\url{https://doi.org/10.18637/jss.v053.i09}}.
		
		\bibitem[James \em{et~al.}(2021)James, Menzies, and
		Bondell]{James2021_geodesicWasserstein}
		James, N.; Menzies, M.; Bondell, H.
		\newblock Understanding spatial propagation using metric geometry with
		application to the spread of {COVID}-19 in the {U}nited {S}tates.
		\newblock {\em Europhys. Lett.} {\bf 2021}, {\em 135},~48004.
		\newblock {\url{https://doi.org/10.1209/0295-5075/ac2752}}.
		
		\bibitem[James \em{et~al.}(2013)James, Witten, Hastie, and Tibshirani]{ISL}
		James, G.; Witten, D.; Hastie, T.; Tibshirani, R.
		\newblock {\em An Introduction to Statistical Learning}; Springer: New York,
		NY, USA, 2013.
		\newblock {\url{https://doi.org/10.1007/978-1-4614-7138-7}}.
		
		\bibitem[IPC(2005)]{IPCC}
		{Intergovernmental Panel on Climate Change}. Carbon Dioxide Capture and Storage. 2005.
		\newblock
		Available online: \url{https://www.ipcc.ch/site/assets/uploads/2018/03/srccs_wholereport-1.pdf}  (accessed on 26 November 2022). 
		
		\bibitem[Kantorovich and Rubinstein(1958)]{Kantorovich}
		Kantorovich, L.V.; Rubinstein, G.
		\newblock On a space of completely additive functions
		\newblock {\em Vestn. Leningr. Univ.} {\bf 1958}, {\em
			13},~52--59.
		
	\end{thebibliography}

\end{adjustwidth}
\end{document}